# ALMA OBSERVATIONS OF THE SUN IN CYCLE 4 AND BEYOND

*Scientific opportunities for the
first regular observations of the Sun
with the
Atacama Large Millimeter/submillimeter Array*

Prepared by the
Solar Simulations for the Atacama Large Millimeter Observatory Network
(SSALMON)

In cooperation with the solar ALMA development studies
"Advanced Solar Observing Techniques"
and
"Solar Research with ALMA"

VERSION 1.2 - March 29th, 2016

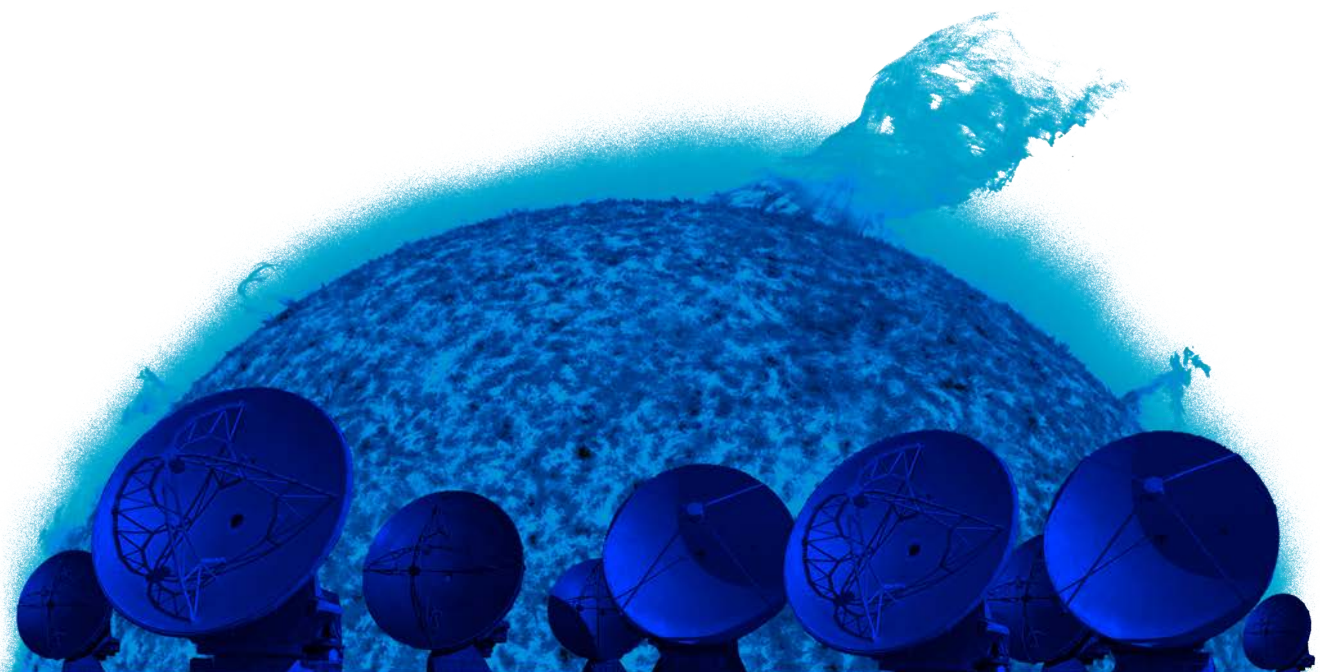

SOLAR SIMULATIONS FOR THE ATACAMA LARGE MILLIMETER OBSERVATORY

# Table of Contents

# 1. ABOUT THIS DOCUMENT

This document was created by the **Solar Simulations for the Atacama Large Millimeter Observatory Network (SSALMON)** in preparation of the first regular observations of the Sun with the **Atacama Large Millimeter/submillimeter Array (ALMA)**, which start in ALMA Cycle 4 in October 2016. The science cases presented here demonstrate that a large number of scientifically highly interesting observations could be made already with the still limited solar observing modes foreseen for Cycle 4 and that ALMA has the potential to make important contributions to answering long-standing scientific questions in solar physics.

During 2015, with the proposal deadline for ALMA Cycle 4 in April 2016 and the *Commissioning and Science Verification* campaign in December 2015 in sight, several of the SSALMON Expert Teams composed strategic documents in which they outlined potential solar observations that could be feasible given the anticipated technical capabilities in Cycle 4. These documents have been combined and supplemented with an analysis, resulting in recommendations for solar observing with ALMA in Cycle 4. In addition, the detailed science cases also demonstrate the scientific priorities of the solar physics community and which capabilities are wanted for the next observing cycles.

The work on this White Paper effort was coordinated in close cooperation with the two international development studies:

- *"Advanced Solar Observing Techniques"* - A project within the North American Study Plan for Development Upgrades of the ALMA
  (PI: T. Bastian, National Radio Astronomy Observatory (NRAO), USA).

- *"Solar Research with ALMA"* - A project carried out at the Czech ARC node of European ALMA Regional Center (EU ARC at Ondrejov, Czech Republic) in the frame of the ESO program "Enhancement of ALMA Capabilities/EoC",
  (PI: Roman Brajsa, Hvar Observatory, Croatia).

We recommend reading the comprehensive overview of potential solar science with ALMA, which was also written by the SSALMON group and complements the document at hand:

*Solar science with the Atacama Large Millimeter/submillimeter Array - A new view of our Sun*
*S. Wedemeyer, T. Bastian, R. Brajsa, H. Hudson, G. Fleishman, M. Loukitcheva, B. Fleck, E. P. Kontar, B. De Pontieu, P. Yagoubov, S. K. Tiwari, R. Soler, J. H. Black, P. Antolin, E. Scullion, S. Gunar, N. Labrosse, H.-G. Ludwig, A. O. Benz, S. M. White, P. Hauschildt, J. G. Doyle, V. M. Nakariakov, T. Ayres, P. Heinzel, M. Karlicky, T. Van Doorsselaere, D. Gary, C. E. Alissandrakis, A. Nindos, S. K. Solanki, L. Rouppe van der Voort, M. Shimojo, Y. Kato, T. Zaqarashvili, E. Perez, C. L., Selhorst, M. Barta (in total 39 authors),* Space Science Reviews (2015, DOI: 10.1007/s11214-015-0229-9)

On behalf of the SSALMON group,

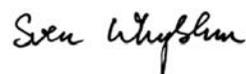

(Sven Wedemeyer)  January 2016

**Document history**

| | |
|---|---|
| October 23rd, 2015 | Strategic Documents finished by SSALMON Expert Teams |
| November 27th, 2015 | First complete version (1.0) of this White Paper |
| January 6th, 2016 | Minor updates in response of the positive decision on solar observing in Cycle 4 |
| March 29th, 2016 | Update of the technical capabilities (in particular, duration of consequent observing) and corresponding adjustments of science case requirements/observing plans. |

## 2. THE SOLAR SIMULATIONS FOR THE ATACAMA LARGE MILLIMETER OBSERVATORY NETWORK (SSALMON)

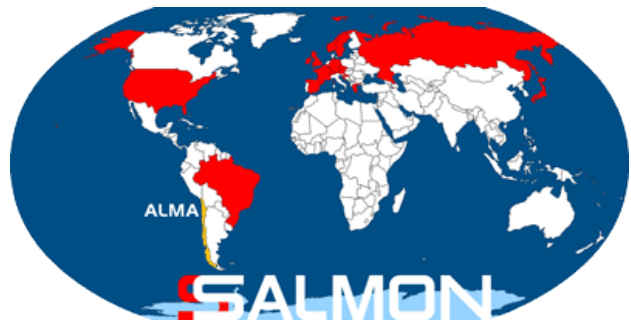

SSALMON is an international scientific network with the aim to coordinate modeling efforts that will help to plan, optimize and analyze solar observations with ALMA. SSALMON was initiated on September 1st, 2014, in connection with on-going solar ALMA development studies. The network has grown rapidly since then and has currently (March 2016) 82 participating scientists from 18 countries from Europe, North and South America and East Asia. This development clearly demonstrates the large interest, which the international solar physics community has in using ALMA. As part of SSALMON 16 Expert Teams focus on different topics ranging from numerical simulations of the solar atmosphere, modelling of ALMA's instrumental properties to scientific applications such as active regions, flares, prominences and wave studies.

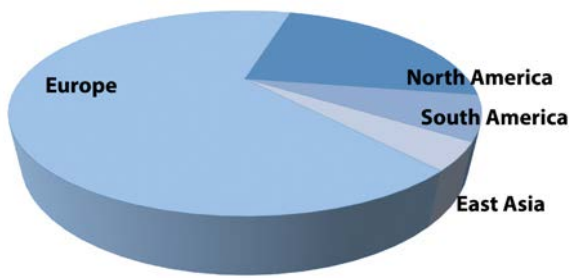

Details and more information about SSALMON can be found on the webpage (http://www.ssalmon.uio.no/) and in an article published in *Advances in Space Research (*Wedemeyer et al. 2015, http://adsabs.harvard.edu/abs/2015arXiv150205601W*).*

**SSALMON Key Goals**

**Key Goal 1 - Raising awareness of science opportunities with ALMA.** Simulations should anticipate and stimulate the solar physics community in thinking about the science opportunities with ALMA. Such simulations demonstrate what could be possibly observed with ALMA and which scientific problems could therefore be addressed in the future. Many scientific topics, which are currently worked on with other instruments, could possibly also be investigated by using ALMA. Many scientists outside the solar radio community are certainly not fully aware of the opportunities and potential of ALMA. Motivating these colleagues to consider ALMA observations for their research could strengthen the visibility and the possible accomplishments of the solar ALMA community.

**Key goal 2 - Clear visibility of solar science within the ALMA community.** Solar observations differ technically from other ALMA observations. Consequently, solar observation runs - apart from initial tests - have been postponed to later observation cycles. SSALMON aims at demonstrating the wider ALMA community that solar observations have a large potential to produce fascinating and important scientific results and should thus receive sufficient ALMA time. Simulations have and can produce illustrative results, which not only demonstrate the feasibility of solar applications but which can also be used to make the scientific potential more visible to the wider ALMA community.

**Key goal 3 - Constraining ALMA observing modes with numerical simulations.** "Observing simulations", i.e., artificial observations based on numerical models of the solar atmosphere are a valuable tool, which will help to constrain and optimize ALMA observing modes. Such simulations show what kinds of observations should ALMA perform in order to usefully test science questions and which of these scientific applications are most promising in view of ALMA's wavelength range and angular resolution. Observing simulations should be used to examine relevant issues like observing bands, antenna configuration, pointing strategy (single pointing, mosaic, on-the-fly mapping), time resolution, etc. This step will involve simulating an observation with the instrument and assessing the results (i.e., determine whether a given simulation adequately addresses the science goals). As a result, recommendations regarding the appropriate observing strategy would be made for particular science goals ("use cases"). In summary, the observing simulations serve two purposes: (1) Demonstration of the feasibility of using ALMA to address the science questions raised. (2) Definition of the specific requirements that ALMA operations must satisfy in order to successfully support such observations.

## 3. CONTRIBUTING AUTHORS

| Name | Affiliation | Contribution |
|---|---|---|
| **S. Wedemeyer** | University of Oslo, Norway | Principal author, lead author of teams J and K |
| **B. Fleck** | ESA Science Operations Department | Lead author of team G |
| **M. Battaglia** | University of Applied Sciences Northwestern Switzerland, Switzerland | Lead author of team H |
| **N. Labrosse** | University of Glasgow, UK | Lead author of team I |
| **G. Fleishman** | New Jersey Institute of Technology, USA | Lead author of team L, team H |
| **H. Hudson** | UC Berkeley, USA; University of Glasgow, UK | Lead author of team P, team H |
| **P. Antolin** | National Astronomical Observatory of Japan (NAOJ), Japan | Teams G, H, I , J, P |
| **C. Alissandrakis** | University of Ioannina, Greece | |
| **T. Ayres** | University of Colorado, USA; Center for Astrophysics & Space Astronomy, USA | Teams J, K |
| **J. Ballester** | Univ. Illes Balears, Spain | Teams G, I |
| **T. Bastian** | National Radio Astronomy Observatory (NRAO), USA | Cycle 4 capabilities, PI of NA development study |
| **J. Black** | Chalmers University of Technology, Dept. of Earth and Space Sciences, Sweden | Spectral lines |
| **A. Benz** | University of Applied Sciences Northwestern Switzerland, Switzerland | Teams H |
| **R. Brajsa** | Hvar Observatory, Croatia | Team I, PI of EU development study |
| **M. Carlsson** | University of Oslo, Norway | Teams J, K |
| **J. Costa** | National Institute for Space Research - INPE, Brazil | Teams H,P |
| **B. DePontieu** | LMSAL, USA | IRIS co-observing |
| **G. Doyle** | Armagh Observatory, UK | Teams G, H, J |
| **G. Gimenez de Castro** | Universidade Presbiteriana Mackenzie, Brazil | Teams H, P |
| **S. Gunár** | Astronomical Institute, Academy of Sciences, Czech Republic | Team I |
| **G. Harper** | University of Colorado at Boulder, USA | Team K |
| **S. Jafarzadeh** | University of Oslo, Norway | Team K |
| **M. Loukitcheva** | Saint-Petersburg State University, Russia | Teams G, J , L, K |
| **V. Nakariakov** | University of Warwick, UK | Teams G, H |
| **R. Oliver** | University of the Balearic Islands, Spain | Teams G, P |
| **B. Schmieder** | Observatoire de Paris Meudon, France | Team I |
| **C. Selhorst** | UNIVAP - University of Vale do Paraíba, Brazil | Team P |
| **M. Shimojo** | National Astronomical Observatory of Japan (NAOJ), Japan | Cycle 4 capabilities |
| **P. Simões** | University of Glasgow, UK | Team H |
| **R. Soler** | University of the Balearic Islands, Spain | Team G |
| **M. Temmer** | University of Graz, Austria | Team I |
| **S. Tiwari** | NASA's MSFC, USA | Team J |
| **T. Van Doorsselaere** | KU Leuven, Belgium | Team G |
| **A. Veronig** | Kanzelhöhe Observatory & Institute of Physics/ University of Graz, Austria | Team H |
| **S. White** | AFRL, USA | Cycle 4 capabilities |
| **P. Yagoubov** | ESO, Germany | Cycle 4 capabilities |
| **T. Zaqarashvili** | Institute of Physics, University of Graz, Austria | Team G |

## 4. THE SUN AS AN EXOTIC AND PROMISING TARGET FOR ALMA

The proximity of the Sun does not only enable life on Earth but it also makes observing the Sun different from most other astronomical objects. Spatially resolved observations with the advanced instrumentation that became available during the recent decades have radically changed our picture of the Sun. We know now that our host star is a highly dynamic object with a complicated structure in its outer layers, which varies spatially and temporally on a large range of scales (see Fig. 4.1). The Sun's **multi-scale nature** makes it a very interesting object of study but also a challenging one. The smallest scale resolved with currently existing solar telescopes is on the order of 0.1 arcseconds at visible wavelengths but the Sun is most likely structured on even smaller, yet unresolved scales.

In general, the temporal and spatial scales correlate in the solar atmosphere so that, apart from explosive events, most larger structures evolve slower than the smallest structures. The small field-of-view of interferometric observations with ALMA, given by the primary beam size of a single ALMA antenna, contains thus a lot of **small-scale, fast evolving structure**. Typical time scales in the chromosphere, which is the layer from where most of the radiation at millimeter wavelengths is emitted, are on the order of a few tens of seconds or even less. Consequently, a high temporal resolution is required or otherwise the small-scale structure is lost, i.e. smeared out when integrating over too long time windows. Using the Earth's rotation during an observation for increasing the u-v coverage, i.e. the spatial scales sampled by the interferometric array, is thus not an option for observations of the Sun. On the other hand, **the Sun is a very bright source at ALMA wavelengths** which allows for snapshot observing with integration times of well below 1 s. Snapshot imaging should be possible for ALMA with its many antennas in a compact configuration, ensuring sufficiently dense sampling in the u-v plane to measure all of the relevant angular scales.

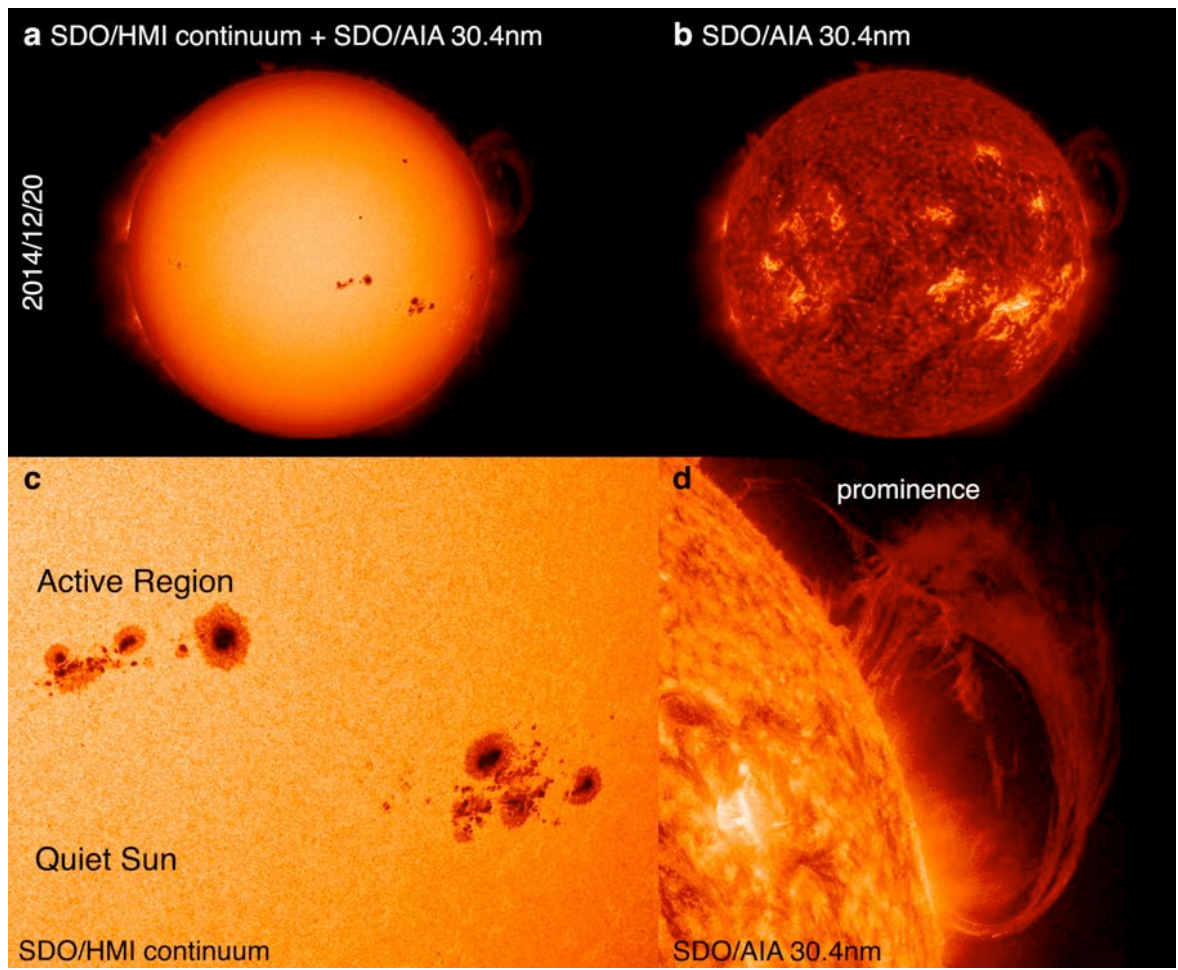

*Figure 4.1:* The Sun on December 20$^{th}$, 2014, as observed with NASA's Solar Dynamics Observatory (SDO). *a)* Continuum image in the visible (617 nm) with the 30.4 nm channel in the background. *b)* 30.4 nm channel mapping a layer above. *c)* Close-up of a on-disk region with Active Regions and surrounding Quiet Sun. *d)* A prominence at the limb. Credits: SDO/NASA

**Forming a new scientific field: High-resolution solar millimeter observations**

The spatial resolution achieved for observations of the Sun at millimeter and radio wavelengths has been much lower than what was possible at shorter wavelengths and generally limited to resolutions ranging from few arcsec (e.g., with interferometers like the Berkeley-Illinois-Maryland Array, BIMA) to ~20 arcsec for single dish observations. Much of the small-scale structure so distinctly visible at visible and ultraviolet (UV) wavelengths could not be observed in the millimeter and radio range due to the too low spatial resolution. In the past, solar radio observations have therefore usually addressed other scientific questions than what was done for based on observations at visible, UV and infrared (IR) wavelengths. The result was a rather small overlap of the solar radio and the high-resolution optical/UV/IR scientific communities. This situation is now changing with ALMA because this interferometric array can achieve a spatial resolution at its shortest wavelengths, which is comparable to what is currently achieved with optical solar telescopes. Consequently, ALMA will be able to investigate those scientific questions that had been beyond reach at millimeter wavelengths before and had been addressed through observations at shorter wavelengths. ALMA is thus also bringing together the scientific community, forming a new field, which could be described as **high-resolution solar millimeter science**. (see Fig. 4.2).

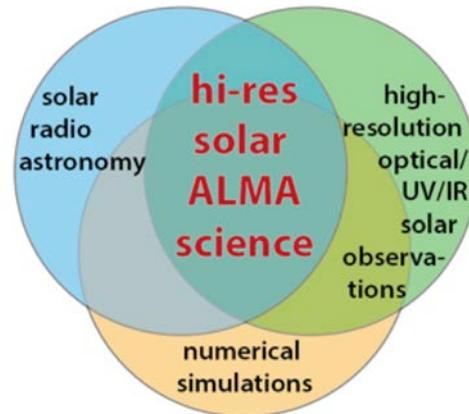

*Figure 4.2:* The high spatial resolution achieved with ALMA brings together formerly rather distinct scientific communities.

**ALMA has unique capabilities, which promise new high-impact scientific results** and progress for answering fundamental long-standing questions in solar physics. As already mentioned before, the radiation observed by ALMA originates mostly from the solar chromosphere. This layer is sandwiched between the photosphere, which is the "surface" of the Sun at visible wavelengths (see Fig. 4.1a,c), and the transition region and corona (see Fig. 4.3). The chromosphere is difficult to observe because not many suitable diagnostics are available and are difficult to interpret due to complicated formation mechanisms and non-linear effects. On the other hand, the chromosphere is a layer of fundamentally important layer because the energy powering the solar corona and the solar wind has to be transported through this enigmatic layer.

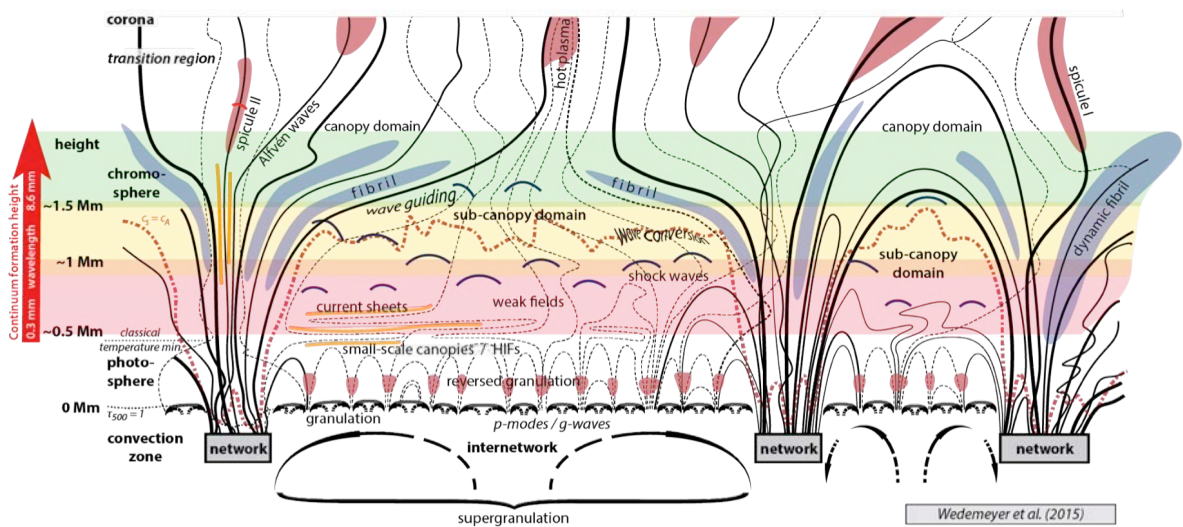

*Figure 4.3:* Schematic height structure of the solar atmosphere in quiet Sun regions ranging from the photosphere to the transition region and corona. The colored areas illustrate the chromospheric layers probed by ALMA at different wavelengths. From Wedemeyer et al. (submitted to SSRv).

The following capabilities are not provided by other instruments, thus making ALMA a special tool for observing the Sun with great scientific potential:

1. The continuum radiation at millimeter wavelengths acts as a **linear thermometer** for the gas in a rather narrow layer in the solar atmosphere.
2. The polarization provides a measure of the longitudinal **magnetic field** component in the same layer in the solar atmosphere.
3. The height of the probed atmospheric layer increases with the selected wavelength, enabling height scans through the solar atmosphere and **tomographic techniques**.

In other words, ALMA should be able to measure the **time-dependent, three-dimensional structure of the solar atmosphere**, providing a complete <u>unprecedented</u> description of the thermal, magnetic and kinetic state of the atmospheric gas.

**Fundamental questions in solar physics to be addressed with ALMA**

Among the many possible solar ALMA science cases, the following topics are fundamental problems in contemporary solar physics. Results based on ALMA observations, which contribute to progressing towards a solution of the problems, would have high impact. Please refer to the review article mentioned in Sect. 1 for more details.

- **Coronal and chromospheric heating.** Since the late 1930ies it is clear that the observed high temperatures in excess of a million degree Kelvin can only be explained by heating processes but it is still not known how exactly this happens. The lack of definite answers is to a good part due to the lack of reliable diagnostics for probing the involved layers in the solar atmosphere. Being a novel tool, ALMA has the potential to substantially contribute to solving the coronal/chromospheric heating problem by probing the 3D thermal structure and dynamics of the solar chromosphere, thus revealing the energy transport in the outer layers. Chromospheric waves and oscillations are important in this context and could already be investigated starting in Cycle 4.

- **Solar flares** are triggered by the rapid reconfiguration of the magnetic field in Active Regions (see Fig. 4.1a), resulting in the violent release of large amounts of energy in the form of radiation and high-energy particles. Major flares occur often in connection with coronal mass ejections (CME), which can enter the interplanetary space and affect Earth. Despite the potential impact of flares, there are still many open questions about the basic physics of flares at all scales including, e.g., particle acceleration mechanisms and the source of the still enigmatic emission component at sub-THz frequencies. ALMA's ability to probe the thermal and magnetic state of flaring regions at high spectral, spatial and temporal resolution promises ground-breaking discoveries.

- **Solar prominences** (a.k.a. filaments when seen on-disk) are large-scale structures comprised of cool and dense gas supported by magnetic fields against gravity (see Fig. 4.1d). Quiescent prominences can remain stable for days to weeks, while active prominences can evolve and erupt on short time scales. Such eruptions can emit gas into the interplanetary space and thus are a source of "space-weather". ALMA is an ideal tool for probing the cool prominences plasma and addressing the many unanswered questions from the formation to the eruption of prominences in unprecedented detail.

## 5. SIMULATIONS AND PREDICTIONS OF SOLAR ALMA SCIENCE

As outlined in the previous section, ALMA enters a new domain in terms of spatial resolution at millimeter wavelengths for solar observations. Consequently, appropriate diagnostic tools and observing strategies have to be developed. While a lot can be learned from existing high-resolution observations at shorter wavelengths, more detailed approaches rely on using state-of-the-art numerical models of the solar atmosphere and corresponding artificial observing, which simulate how ALMA would and should observe various solar targets. Detailed numerical models allow for testing different instrumental set-ups, e.g., different array configurations, receiver bands, and effect on the resulting data sets. Artificial observations of numerical models at millimeter wavelengths are therefore an important tool for planning, optimizing and interpreting future solar observations with ALMA and thus the central objective of SSALMON. Results from these simulations will be used to further refine the science cases presented in this document, which will be accordingly updated during the time until Cycle 4 begins.

The continuous comparison of numerical simulations and observations has driven much progress in solar physics during the past decades and will continue doing so (see Fig. 5.1 for an illustration of this strategy). The comparisons reveal which physical processes are still missing or not described correctly in the models, thus leading to true understanding of how the solar atmosphere works. On the other hand, numerical simulations can predict yet unobserved effects and processes, which can lead to discoveries.

This approach works also brilliantly for ALMA observations as can be seen from examples in this document. For instance, three-dimensional models of Quiet Sun regions have been used to calculate synthetic brightness temperature maps, which can be used to adjust the required observing set-up and, since Quiet Sun regions are relatively well understood, possibly for calibration in the future.

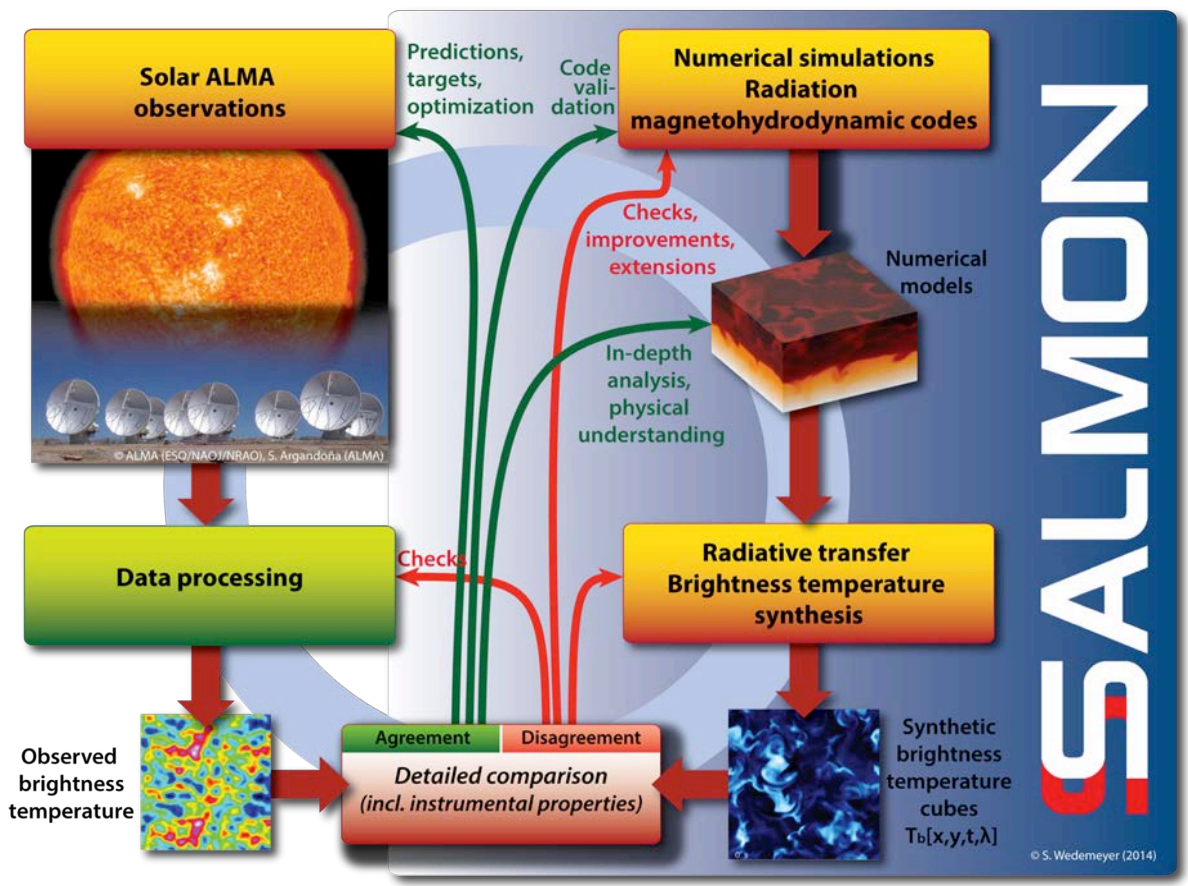

*Figure 5.1.* Detailed comparisons of ALMA observations (left) with numerical models (right) enable us to develop observing strategies for ALMA, 45 to plan, optimize, and interpret observations, and to demonstrate that potentially important scientific results can be expected from solar ALMA observations. From Wedemeyer et al. (2015, AdSR 56, 2679).

# 6. SUMMARY OF PROPOSED SCIENCE CASES FOR CYCLE 4

The technical capabilities of the solar observing modes that will most likely be offered in ALMA Cycle 4 (see Sect. 8) are still limited but already allow for a large number of highly interesting studies. On the other hand, some SSALMON expert teams, like e.g. Team D (*Spectral lines in the millimeter range as new diagnostic tools*) and Team F (*Magnetic field measurements*) rely on capabilities, which are not yet offered, and will start their work once decided if and when these capabilities can be expected. The remaining SSALMON expert teams have suggested the following topics for Cycle 4:

- **Quiet Sun regions (Team K):** The brightness temperature distribution of the chromosphere and its variation with height is an important test case for comparisons with solar atmosphere models and thus an essential step towards understanding this important interface layer of the Sun. A sound understanding of the brightness temperature distribution would allow for self-calibrations of solar ALMA observations.

- **Active regions (Team L):** ALMA observations of active regions and sunspots will result in important constraints on the thermal and magnetic field structure of these regions as it can be derived from the impact on the brightness temperature and from oscillations and waves.

- **Solar flares (Team H)** are very energetic events during which large amounts of energy are released in the form of radiation across all wavelengths, incl. the range accessible to ALMA, and high-energy particles. Fundamental but yet unanswered questions about the nature of flares and their potential impact on Earth makes solar flare studies a very active field of research, to which ALMA can significantly contribute. As a first step, the temporal and spatial evolution of solar flares at ALMA wavelengths should be investigated. The spectral information anticipated for Cycle 4 already gives new insights into the energy spectrum of flares.

- **Prominences (Team I):** ALMA can help to answer important open questions about prominences (off-limb) and filaments (on-disk). It includes detailed observations of the thermal structure of prominence bodies and their spatial (threaded) fine-structure. Waves and oscillations may provide essential information, incl. constraints on the magnetic field structure of prominences.

- **Oscillations and waves (Team G)** are interesting for several reasons. They are important for the structure and dynamics of the solar atmosphere, they provide potential channels for energy into the upper layers, and provide means to probe the structure of the solar atmosphere. Given that ALMA serves as a linear thermometer of the chromospheric gas at high temporal, spatial, and spectral resolution, observations of oscillations and waves will provide important new constraints on the structure of different regions on the Sun *(which is the reason why many science cases of team G overlap with other teams)*.

- **Chromospheric and coronal heating (Team J):** Apart from the role of waves for the heating of the chromosphere, the connection between the chromospheric and coronal plasma and the involved energy transfer is an important topic that can be addressed with ALMA. High-cadence observation sequences will allow to search general heating signatures and also to focus on specific known processes.

- **Limb-brightening studies (Team P):** Observations of the variation of the brightness temperature as function of the distance from the solar limb would allow for deriving the mean gas temperature stratification of the solar atmosphere and would serve as an important first step towards a detailed interpretation of solar observations at ALMA wavelengths. Observations at the solar limb are technically challenging because the side-lobes of the ALMA point spread function would result in large contributions to the signal from the bright solar disk, potentially drowning the off-limb signal.

More details and science cases are given in the corresponding team sections in the second half of this document.

There is a natural **overlap** between some of the proposed science cases, mostly due to the choice of the solar targets (cross-references are given at the end of each science case). Some of the proposed observations would therefore produce data sets that are of interest for more than one team, which would analyse different aspects of the data, thus making the best possible use of it. The latter is important in view of the limited observing time. For instance, ALMA would not wait for a flare but rather follow an active region, which is likely to produce flares. This way, the same observation is again interesting for a number of different, complementary aspects. Consequently, Active Regions are not only the target for Team L but for many more science cases.

**The solar atmosphere is a very dynamic and spatially intermittent phenomenon.** Observing the temporal evolution of a solar target at high spatial resolution is therefore essential for understanding the processes at work in the solar atmosphere. Consequently, most observations proposed here require time sequences of interferometric brightness temperature maps. The needed integration time is anticipated to be below 1s because ALMA is very sensitive and the Sun a bright radio source. The achievable temporal resolution of an observation is rather constrained by the required field of view (FOV). ALMA's FOV for interferometric observations is quite small and actually too small for the largest structures such as entire active regions or prominences. The effective FOV can be increased by mosaicking, i.e., rapidly changing the pointing and combining the maps afterwards. The time for completing a mosaic, which is then the cadence of the observation, increases with the number of pointings (plus time for intermediate calibration). For 150 pointings, which might be the maximum possible in Cycle 4, the cadence would decrease to 40min. The optimum compromise between spatial and temporal resolution depends on the requirements of the individual science cases. Sequences at a single, fixed pointing with a high cadence of 10s or even down to the limit of 2s for Cycle 4 are suitable for observing a number of dynamic small-scale phenomena whereas temporal resolution can be sacrificed for observing large-scale structures like prominences. A typical interferometric data set would in any case consist of a sequence of 1-2 hours with many maps with short integration time and, in Cycle 4, for 4 sub-bands with up to 128 channels each. The provided spectral information is already of great value as the sampled height in the solar atmosphere depends on the wavelength and thus enables a first view of the atmospheric stratification.

**Fast-scanning** with a single (or up to four) Total Power antennas is interesting for solar observations. Full-disk scans of the Sun can be achieved with a cadence of a few minutes and provide important context information for simultaneous interferometric observations, which are confined to a small region on the Sun only. In addition, "single-dish" observations can provide interesting information about the large-scale structure of the solar atmosphere and, when scanning the limb of the solar disk, its stratification.

While ALMA data sets are of very high scientific value for themselves, their potential is further enhanced as part of **coordinated observing campaigns** with other instruments that provide complementary diagnostic information of the same layer as probed by ALMA and also important context information of the layers below and above. The latter is essential because the solar atmosphere is a compound of coupled domains of which the chromosphere, as observed with ALMA, is an integral part. Co-ordinated observing campaigns with ground-based and space-borne telescopes are routinely carried out and have become a standard procedure in solar physics. More details about instruments, which are anticipated to co-observe with ALMA in Cycle 4 (and beyond), are given in Sect. 9.

**Summary of requested observing set-ups**

The tables at the end of this section show that many science cases have similar technical requirements. The most frequent targets are **Active Regions** as they can be used for a multitude of scientific goals incl. flares and oscillations. Given the limited capabilities foreseen for Cycle 4, in particular due to the long calibration breaks interrupting the science observations, the majority of science cases requests a **interferometric high-cadence sit-and-stare set-up**, i.e. following a pre-selected solar target (accounting for the solar rotation) and producing interferometric observations with a time resolution of 2 s. In most cases, first one run in band 3 and then one run in band 6 for the same target is desirable. Most science cases do not request a particular array configuration as long as it is compact (C40-1/2/3). In addition, **small mosaics** with 6x6 pointings and an overall cadence of 8-10min are another frequently requested set-up. Please note that also the remaining set-ups listed in the table are of high scientific interest.

**Science case overview**

In total, the expert teams have proposed 27 science cases, which are compared below:

| Science cases | G1 | G2 | G3 | G4 | G5 | G6 | G7 | G8 | G9 | H1 | H2 | H3 | I1 | I2 | J1 | J2 | J3 | J4 | K1 | K2 | K3 | L1 | L2 | L3 | L4 | P1 | P2 |
|---|---|---|---|---|---|---|---|---|---|---|---|---|---|---|---|---|---|---|---|---|---|---|---|---|---|---|---|
| **Target** | | | | | | | | | | | | | | | | | | | | | | | | | | | |
| Disk center | | | | | | | | | | | | | | | | | | | | | | | | | | | |
| On disk | | | | | | | | | | | | | | | | | | | | | | | | | | | |
| Off/ close to limb | | | | | | | | | | | | | | | | | | | | | | | | | | | |
| Quiet Sun | | | | | | | | | | | | | | | | | | | | | | | | | | | |
| Magnet. network | | | | | | | | | | | | | | | | | | | | | | | | | | | |
| Enhanced network/ plage | | | | | | | | | | | | | | | | | | | | | | | | | | | |
| Active Region/ sunspot | | | | | | | | | | | | | | | | | | | | | | | | | | | |
| Pore/non-flaring sunspot | | | | | | | | | | | | | | | | | | | | | | | | | | | |
| Prominence/ filament | | | | | | | | | | | | | | | | | | | | | | | | | | | |
| **Interferometry - Array configuration** | | | | | | | | | | | | | | | | | | | | | | | | | | | |
| C40-1 | | | | | | | | | | | | | | | | | | | | | | | | | | | |
| C40-2 | | | | | | | | | | | | | | | | | | | | | | | | | | | |
| C40-3 | | | | | | | | | | | | | | | | | | | | | | | | | | | |
| **Interferometry - pointing** | | | | | | | | | | | | | | | | | | | | | | | | | | | |
| single | | | | | | | | | | | | | | | | | | | | | | | | | | | |
| mosaic, 36-40 p. | | | | | | | | | | | | | | | | | | | | | | | | | | | |
| mosaic 150 p. | | | | | | | | | | | | | | | | | | | | | | | | | | | |
| point. sequence | | | | | | | | | | | | | | | | | | | | | | | | | | | |
| **Interferometry - receiver bands** | | | | | | | | | | | | | | | | | | | | | | | | | | | |
| 3 | | | | | | | | | | | | | | | | | | | | | | | | | | | |
| 6 | | | | | | | | | | | | | | | | | | | | | | | | | | | |
| both bands | | | | | | | | | | | | | | | | | | | | | | | | | | | |
| **Interferometry - temporal resolution (cadence)** | | | | | | | | | | | | | | | | | | | | | | | | | | | |
| *(1 s)* | | | | | | | | | | | | | | | | | | | | | | | | | | | |
| 2 s | | | | | | | | | | | | | | | | | | | | | | | | | | | |
| 10 s | | | | | | | | | | | | | | | | | | | | | | | | | | | |
| 20 s | | | | | | | | | | | | | | | | | | | | | | | | | | | |
| 60 s | | | | | | | | | | | | | | | | | | | | | | | | | | | |
| ~10 min | | | | | | | | | | | | | | | | | | | | | | | | | | | |
| 40 min | | | | | | | | | | | | | | | | | | | | | | | | | | | |
| **Interferometry - duration** | | | | | | | | | | | | | | | | | | | | | | | | | | | |
| > 120 min | | | | | | | | | | | | | | | | | | | | | | | | | | | |
| 60 min | | | | | | | | | | | | | | | | | | | | | | | | | | | |
| 30 min | | | | | | | | | | | | | | | | | | | | | | | | | | | |
| 5-10 min | | | | | | | | | | | | | | | | | | | | | | | | | | | |
| **Science cases** | G1 | G2 | G3 | G4 | G5 | G6 | G7 | G8 | G9 | H1 | H2 | H3 | I1 | I2 | J1 | J2 | J3 | J4 | K1 | K2 | K3 | L1 | L2 | L3 | L4 | P1 | P2 |

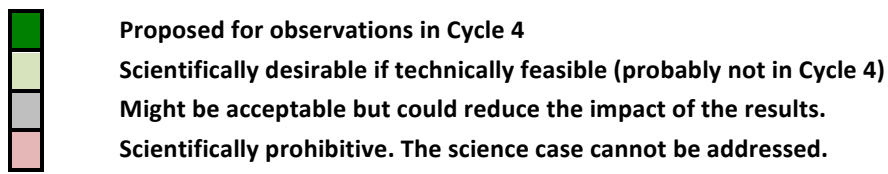

- ■ **Proposed for observations in Cycle 4**
- ■ **Scientifically desirable if technically feasible (probably not in Cycle 4)**
- ■ **Might be acceptable but could reduce the impact of the results.**
- ■ **Scientifically prohibitive. The science case cannot be addressed.**

*Table 6.1: Comparison of the observational requirements for all proposed Cycle 4 science cases.*

### Suggested standard observation set-ups

Most of the proposed observations for Cycle 4 could be carried with the setups listed below.

| INTERFEROMETRIC IMAGING | | | | | | |
|---|---|---|---|---|---|---|
| **ID** | INT-1 | INT-2 | INT-3 | INT-4 | INT-5 | INT-6 |
| **Name** | FAST | MODERATE | SMALL MOSAIC | LARGE MOSAIC | MULTI | SWEEP |
| **Description** | high-cadence single-point sit-and-stare | medium-cadence single-point sit-and-stare | mosaic with few pointings, moderate cadence | mosaic with max. pointings and low cadence | single-point sit-and-stare, prescribed sequence of pointings | prescribed sequence, 5 pointings per block |
| **Pointings** | 1 | 1 (K1: 10 p. changes | 6 x 6 (or up to 40 p.) | 150 | 1 per 10 min science block | 10 per 10min science block |
| **Temporal resolution/cadence** | 2 s *per map* | 10 s *per map* | 8-10 min *per map* | ~40 min *per map* | 2 s - 10 s *per map* | 2s - 60 s *per map* |
| **Duration** | **60 min (default)** *Many science cases request shorter sequences but could use INT-1 data, too.* | **120 min** (J4), otherwise **60 min (default)** *Many science cases request shorter sequences but could use INT-2 data, too.* | min. 1 hour | ~ 80min *(3 full mosaics plus calibration)* | max. 91 min for 7 pointings *(10 min per pointing plus 3 min calibration)* | 65 min for 50 pointings *(1 min per pointing plus 5 x 3 min calibration)* |
| **Receiver band** *(as requested for the science cases)* | • **only 3:** H2, J1<br>• **only 6:** G3, G6, G7, G8<br>• **first 3, then 6:** default. | • **only 3:** J3<br>• **only 6:** G9<br>• **first 3, then 6:** J4, K1-3 | • **3 only:** H3<br>• **6 only:** H1<br>• **first 3, then 6:** L1-L3, I1 | • **first 3, then 6:** I2 | • **first 3, then 6:** K1-2 | • **first 3, then 6** |
| **Target** | Versatile *(see comments in table below)*. | • Active Region (G9)<br>• Sunspot (J3)<br>• Quiet Sun (K1-3) | • Active Region (H1,H3) | Prominence/ filament | Quiet Sun | Disk-center to off-limb, sequence of positions (out to 1.1 $R_{sun}$) |
| **Science cases** | G1-G8, H2, J1, J2, L4 | G9, J3, J4, K3 | H1, H3, I1,I2, L1-L3 | I2 | K1-2 | P1, P2 |
| **Number** | 12 | 4 | 7 | 1 | 2 | 2 |

| FAST SINGLE-DISH SCANNING | | |
|---|---|---|
| **ID** | SD-1 | SD-2 |
| **FOV** | 2400" x 2400" | 3000" x 3000" |
| **Temporal resolution** | ~7-10 min | ~7-10 min |
| **Science cases** | Fast single-dish scanning is a useful component for basically all science cases because it provides valuable context data, in particular when co-observing with other instruments. | |

| Targets for INT-1 | |
|---|---|
| G1: Quiet Sun | G6: chrom. fibril near non-flar. AR |
| G2: Intense magn. field, plage, network, AR | G7: filament near disk center |
| G3, H2, J2: AR | G8: footpoints of AR arcade |
| G4: AR plage | J1: AR/network, loop footpoints |
| G5: chrom. network disk center | J2: Quiet Sun |

*Table 6.2: Suggested standard interferometric and fast-scanning observing set-ups, which would cover all so far proposed solar science cases for Cycle 4.*

# 7. RECOMMENDATIONS FOR THE FUTURE DEVELOPMENT OF SOLAR OBSERVING WITH ALMA

Based on the science cases presented in this document and the general overview of potential solar science with ALMA mentioned in Sect. 1 (Wedemeyer et al. 2015, SSRv), we make the following recommendations for the future development of solar observing modes for Cycle 5 and beyond, ordered by priority. A summary of the items with highest priority is given at the end of this section.

1. **Calibration and continuous observing.** The observations anticipated for Cycle 4 are to be split into science blocks of 10 min with breaks for calibration lasting for 3 min. The resulting data sequences would have large gaps, which is problematic for most solar science cases. An event that is detected in an science observation block might have ended during a calibration break, resulting in only partial observations of such events with limited scientific value. Such gaps are in particular problematic for wave and oscillation studies for which uninterrupted sequences of long duration are ultimately needed. Reducing the frequency and length of intermediate calibration breaks is therefore of high importance and will be tested during the CSV campaign in December 2015. Self-calibration using the well-known characteristics of solar brightness temperature distributions, especially in quiet Sun regions, might offer a solution. Experience gained during Cycle 4 could be decisive in this respect. Another solution is the usage of sub-arrays although the resulting loss in spatial information/resolution is acceptable for some science cases but rather undesirable for other cases. See point 5 for more details on the usage of sub-arrays.

2. **Complementary single-dish fast** scanning is an essential component of any solar observing mode and is therefore requested for Cycle 4. However, more flexible setups of the fast scanning are ultimately necessary in the future. Next to full-disk scans, scans of smaller field of views at then higher cadence have a very high scientific potential. For instance, scans of regions just big enough to cover an active region would increase the probability to observe solar flares while still achieving a reasonable overall cadence. In addition, simultaneous scanning with several TP antennas with a fixed pointing offset could be tested. It should emphasized that fast scanning is not only producing maps that are themselves of scientific interest, but they are also required in combination with interferometric data, in particular for mosaics, for recovering flux on the largest angular scales.

3. **Polarization capabilities** are ultimately needed for addressing a large range of topics because magnetic fields are an essential building block of the solar atmosphere. Utilizing polarization for chromospheric magnetic field measurements has therefore a very high priority and should be offered as soon as technically possible. Simulations for a wavelength of 3 mm predict circular polarization on the order of 3-4% for active regions, i.e. regions with strong magnetic fields, and about ~0.5% for quiet Sun regions, i.e. regions with weak magnetic fields. An accuracy of ~0.1% should therefore enable magnetic field measurements for all considered solar targets. ALMA observations may facilitate, for the first time, the direct determination of the 3D magnetic topology of the solar chromosphere because different wavelengths probe slightly different atmospheric heights in the solar atmosphere. The anticipated high-impact results make it highly desirable to have the polarization capability as early as possible.

4. **Fast receiver band switching** would be a feature with enormous scientific potential. The probed height in the solar atmosphere increases with the selected wavelength so that quickly switching between receiver bands would allow for probing different layers of the atmosphere in quick succession, ultimately leading to tomographic techniques. The resulting unprecedented data sets could describe the time-dependent 3D structure of the chromospheric plasma, which has the potential to revolutionize our understanding of this important interface layer. However, the time for switching between receivers might be as long as 30 min in Cycle 4 and would have to be reduced to less than 1 min given the short dynamical time scales on which the solar chromosphere develops. A temporary solution for multi-band observations is the usage of sub-arrays (see below).

5. **High spectral resolution:** The diagnostic potential of radio recombination lines and molecular lines for the solar atmosphere is still largely unexplored. Numerical models and observations at other wavelengths imply that suitable lines should exist at ALMA wavelengths although the line widths and depths are not easy to predict. Instead, experimental observations are required which would require making use of the in principle flexible setup of the receiver bands and a much larger number of channels than available in Cycle 4.

6. **Sub-arrays:** Beyond Cycle 4, it might be possible to split the 12m-Array into two or more sub-arrays. This capability might be tested in December 2015 and possibly offered in Cycle 5. Sub-arrays would enable two major observing modes for mitigating problems with fast receiver band switching and continuous observing.

    - **Simultaneous multi-band observations** could be achieved by choosing different receiver bands for the sub-arrays like shown in the example below. This option would provide truly simultaneous multi-band observations and thus simultaneous mapping different height ranges in the solar chromosphere at the same time. However, the science observations would still be interrupted by intermediate calibration breaks, which should be as short as possible. It would be reasonable to select antennas for the sub-arrays in such a way that the u-v coverage is as similar as possible in the different bands.

    |  | *30 min* | *10 min* | *3 min* | *10 min* | *3 min* |
    |---|---|---|---|---|---|
    | **Sub-array 1** | initial calibration | **science, band 3** | calibration | **science, band 3** | calibration |
    | **Sub-array 2** | initial calibration | **science, band 6** | calibration | **science, band 6** | calibration |

    - **Continuous single-band observing** could be achieved by alternating between two sub-arrays, which observe the same target in the same receiver band with the same temporal resolution (see example below). While the first sub-array carries out scientific observations, the second sub-array is calibrated and vice versa. The result is a continuous sequence of scientific observations in the selected receiver band. The u-v coverage of the two sub-arrays should be as similar as possible.

    |  | *30 min* | *10 min* | *10 min* | *10 min* | *10 min* |
    |---|---|---|---|---|---|
    | **Sub-array 1** | initial calibration | **science, band 3** | calibration (3 min) | **science, band 3** | calibration (3 min) |
    | **Sub-array 2** | initial calibration | wait | **science, band 3** | calibration (3 min) | **science, band 3** |

    It is important to note that the sub-arrays would have fewer antennas and thus fewer baselines, which would affect the information content of the reconstructed images, corresponding to a reduced spatial resolution. Nevertheless, this option is interesting for science cases for which spatial resolution can be sacrificed.

7. **High temporal resolution:** A cadence of 2 s, as it will be offered most likely in Cycle 4, is already sufficient for many scientific applications. However, for observations of high-frequency waves, which have important implications for the heating of the solar atmosphere, a cadence of 1s or less would be interesting. Furthermore, there are science cases, which would even benefit from a temporal resolution of down to a few milliseconds. Chromospheric and coronal heating is a central question in solar physics since its discovery in the late 1930's and high-cadence ALMA observations have the potential to significantly contribute to a solution of this problem with implications for stellar atmospheres in general.
Based on experience to be gained during the CSV campaign in December 2015 and regular Cycle 4 observations, the optimum integration time and thus the overall cadence has to be determined, which will depend on the complexity of the target structure and the intended scientific aim. The optimum compromise between spatial and temporal resolution has to be chosen according to the requirements and priorities of the individual science cases.

8. **Off-limb scanning:** Observations at the limb during cycle 4 and above will allow to determine the influence of the large brightness gradient on interferometric maps, as well as the extension of the side-lobes influence. This is of primary importance for off-limb studies of prominences, coronal rain and, more generally, solar atmospheric structure, where the dynamic range for interesting but faint features can be observed.

9. **Mosaicking:** The field of view (FOV) is determined by the primary beam size of an 12-m ALMA antenna and is comparatively small, especially for the shortest wavelengths. Larger FOVs can be achieved with mosaics, i.e., by stitching together maps from several consequent pointings. Mosaicking is always a compromise between FOV size and temporal resolution. The larger the number of pointings, the longer it takes to complete one mosaic, the lower the temporal resolution. In the extreme case of 150 pointings as foreseen for Cycle 4, the completion of one mosaic would take 25 min, including the time for intermediate calibration. For the future it is highly desirable to optimize mosaics by (1) minimizing the time for each pointing and (2) maximizing the step size between the pointings. It might be practical to predefine standard mosaics with different sizes from small to large and corresponding temporal resolution.

10. **Multi-frequency synthesis** is not a technical capability but rather a technique using ALMA data sets with many simultaneous spectral channels. The technique is mentioned here because of its large potential for solar observing. The spectral channels of a receiver band sample slightly different components in the u-v plane (i.e., the spatial Fourier components sampled by the interferometer). The spectral information can thus be used to improved the sampling of the u-v plane ("u-v-coverage") and thus enhancing the effective spatial resolution of the reconstructed interferometric images, of course then with accordingly reduced spectral resolution.

11. **Coordinated observations with other telescopes:** In addition to the instruments mentioned in detail in Sect. 9, new space-borne and ground-based observatories will become available for co-observations with ALMA. For instance, the 4-m **Daniel K. Inouye Telescope** (DKIST) on Hawaii is expected to see first light in 2019, while **Solar Orbiter** and possibly the 4-**m European Solar Telescope** will join in the 2020s. It seems highly advisable to create a common strategy and tools (e.g., a web interface) to assist the coordination of all participating instruments. Carrying out solar ALMA observations as campaigns (about a week long) would make it possible for a (solar) scientist at the ALMA site to optimize the schedule for the requested targets throughout the campaign since not all target types are always visible at that time. That is in particular the case for active regions and prominences/filaments.
Solar ALMA campaigns would be carried out ideally in a compact configuration. Currently, it is assumed that suitable compact configurations could be offered in December (or even January and March) but the observing conditions are challenging for some observatories on the Northern hemisphere during these months (e.g., on the Canary Islands). For exploiting the potential of the affected observatories, it would therefore be desirable to have (additional) solar ALMA campaign scheduled in the period April - September if the configuration schedule permits.

| Concluding statement on high-priority development items |
|---|

For fully exploiting ALMA's potential for observing the Sun and contributing fundamental questions with implications for our understanding stars in general, the development of more **continuous observing modes** has the highest priority. The limit of 10 min blocks, as foreseen for Cycle 4, limits most of the here presented and future science cases.
Next, ALMA's **polarization capabilities** would open up a wide range of further scientific opportunities because detailed measurements of the magnetic field in the solar atmosphere are ultimately important and consequently a hot topic in contemporary solar physics with great potential.
**Fast switching of receiver bands** would allow to develop tomographic techniques for ultimately producing unprecedented data sets of the full 3D time-dependent thermal and magnetic structure of the solar chromosphere, which has the potential to revolutionize solar physics.
Further enhancing the **spectral capabilities** towards thousands of spectral channels per band (instead of 4 x 128 as foreseen for Cycle 4) will lead to discoveries and possibly the development of novel and complementary diagnostics of the plasma state in the chromosphere and possibly the corona of the Sun.

## 8. ANTICIPATED CAPABILITIES OF ALMA FOR CYCLE 4

Please note that this document had been written before it was finally decided which solar observing modes will be offered in Cycle 4. The following list was composed in order to enable the SSALMON expert teams evaluating which scientific observations might be possible in Cycle 4. At the time of writing, the technical capabilities listed below were believed to be likely to be offered for solar observing in Cycle 4.

We would like to thank the many devoted colleagues working on these preliminary recommended capabilities as part of CSV/EOC activities done by the solar development teams of the EA/NA/EU-ARCs and JAO. Special thanks go to M. Shimojo, S. White, T. Bastian, A. Remijan, P. Yagoubov and R. Hills.

**1. Array configuration and schedule**

- We anticipate that for Cycle 4, solar observing will be offered both in **Interferometry + Total Power** ("single-dish" fast scanning) mode. *These capabilities were demonstrated in 2014-15 through the NA ALMA development study and the ESO program study.*
- For interferometric observations, a heterogeneous array will be used, which combines >10 7m- and >40 12m-antennas. The three most compact configurations will be used (C40-1,-2,-3).

  - **C40-1**: Longest baseline 150 m
    ⇒ 3.4"@100 GHz, 1.5"@230 GHz
    - Mar. 16 2016 - Apr. 5 2017 (17 days), best observing 8-22h LST[*]
  - **C40-2**: Longest baseline 250 m
    ⇒ 1.8"@100 GHz, 0.8"@230 GHz
    - Jan. 13 2017 - Jan. 31 2017 (9 days), best observing 4-17h LST
  - **C40-3**: Longest baseline 500 m
    ⇒ 1.2"@100 GHz, 0.5"@230 GHz
    - Dec. 23 2016 - Jan. 12 2017 (11 days), best observing 3-15h LST
    - Apr. 6 2017 - Apr. 26 2017 (11 days), best observing 9-23 LST

    [*] *LST = Local Sidereal Time*

- The Sun can only be observed at an elevation of >40° in order to avoid the antennas shadowing each other when in compact configuration. This constraint limits the observing time as follows:
  - (Southern) summer (>December) [Max]: 13:00-20:00UT
  - Equinox (March + September): 13:30-16:30UT
  - (Southern) winter (>May) [Min]: 15:30-17:30UT
- The array configuration is changing at all times. The array configurations that are suitable for solar observations will be formed in Chilean summer (December - April), except for February (closed for maintenance). See list above.

**2. Total Power Observations**

- In addition to interferometry, up to 3 Total Power (TP) antennas may be available for fast scanning of the whole solar disk
- Only full-disk fields are allowed. The actual duration of a complete scan is about 7 minutes for Band 3 and 10 minutes for Band 6.
- The scan pattern has to be exactly the same for all antennas; they cannot point at different target regions.

**3. Receiver bands / correlator set-up**

- Only band 3 (84 - 116 GHz, 3.6-2.6mm) and band 6 (211 - 275 GHz, 1.4-1.1mm) will be available.
- Both bands provide 4 sub-bands each. The frequencies are fixed (see the table below).
- Time Domain Mode (TDM) only.
- For (single-dish) TP mode, 128 channels are averaged in each sub-band thus resulting in 4 simultaneous avg. channels.
- In interferometric mode, each sub-band has 128 equally space channels. The edges of a spectral window (here in TDM) cannot be used for science, leaving the middle 100-110 channels for each sub-band (simultaneously).
- The spectral resolution in each sub-band is 2 GHz / 128ch = 15 MHz (➞ ~45 km/s @100GHz, ~20 km/s @230GHz
- **The (fast) band cycling function is not yet possible.** Each Science Goal (SG, see point 6) is carried out in **either band 3 or band 6**. Changing receiver bands can only happen between the SG blocks with a gap of possibly up to 30 min due to calibration even though receiver switching alone could be done in 30-60s (plus time for calibration). The times for switching and calibration may potentially be reduced substantially in the future (> Cycle 4). See point 6 for a possible remedy.

### 4. Polarization:

- No full polarization capability in Cycle 4. Only XX and DUAL offered in the Observing Tool (plus indirect observations of magnetic field structures, best by co-observing with other instruments).

### 5. Field of View (FOV) and temporal resolution

- FOV for **single pointing**:
  ~58" @Band 3, ~25" @Band 6
- Temporal resolution (cadence, time per pointing) = 2 seconds.
- Depending on the scientific objective on the complexity of the target region, it could be desirable and possible to use longer durations for the final image reconstruction but then with accordingly lower time resolution.
- The interferometric mapping may allow for **mosaicking** with up to 150 pointings. The number of pointings per mosaic can be chosen freely.
- The step size of mosaicking observations is will most likely be 1/4 to 1/2 beam size. Experiments in the CSV campaign in 12/2015 may provide more constraints on the optimal step size.
- It takes ~7.5s to complete 1 pointing (incl. time for pointing change).
- About 80 pointings per 10 min observing can realistically be achieved. Mosaicking with more than 80 pointings will require breaks for calibration.

| BAND 3 - (Single-dish) Total Power mode (output: 4 x 1 (avg,) frequency channel) | |
|---|---|
| sub-band 1 | centered at 93 GHz, 128 channels equally spaced over the window [92 GHz, 94 GHz]; output: average over all 128 channels (i.e. ONE avg. frequency) |
| sub-band 2 | centered at 95 GHz, 128 channels equally spaced over the window [94 GHz, 96 GHz]; output: average over all 128 channels (i.e. ONE avg. frequency) |
| sub-band 3 | centered at 105 GHz, 128 channels equally spaced over the window [104 GHz, 106 GHz]; output: average over all 128 channels (i.e. ONE avg. frequency) |
| sub-band 4 | centered at 107 GHz, 128 channels equally spaced over the window[106 GHZ, 108 GHz]; output: average over all 128 channels (i.e. ONE avg. frequency) |
| **BAND 3 - Interferometric mode** (output 4 x ~100 simultaneous frequency channels) | |
| sub-band 1 | centered at 93 GHz, 128 channels equally spaced over the window [92 GHz, 94 GHz]; output: ALL 128 CHANNELS SIMULTANEOUSLY (middle 100-110 useable) |
| sub-band 2 | centered at 95 GHz, 128 channels equally spaced over the window [94 GHz, 96 GHz]; output: ALL 128 CHANNELS SIMULTANEOUSLY (middle 100-110 useable) |
| sub-band 3 | centered at 105 GHz, 128 channels equally spaced over the window [104 GHz, 106 GHz]; output: ALL 128 CHANNELS SIMULTANEOUSLY (middle 100-110 useable) |
| sub-band 4 | centered at 107 GHz, 128 channels equally spaced over the window[106 GHZ, 108 GHz]; output: ALL 128 CHANNELS SIMULTANEOUSLY (middle 100-110 useable) |
| **BAND 6 - (Single-dish) Total Power mode** (output: 4 x 1 (avg,) frequency channel) | |
| sub-band 1 | centered at 230 GHz, 128 channels equally spaced over the window [229 GHz, 231 GHz]; output: average over all 128 channels (i.e. ONE avg. frequency) |
| sub-band 2 | centered at 232 GHz, 128 channels equally spaced over the window [231 GHz, 233 GHz]; output: average over all 128 channels (i.e. ONE avg. frequency) |
| sub-band 3 | centered at 246 GHz, 128 channels equally spaced over the window [245 GHz, 247 GHz]; output: average over all 128 channels (i.e. ONE avg. frequency) |
| sub-band 4 | centered at 248 GHz, 128 channels equally spaced over the window[247 GHZ, 249 GHz]; output: average over all 128 channels (i.e. ONE avg. frequency) |
| **BAND 6 - Interferometric mode** (output 4 x ~100 simultaneous frequency channels) | |
| sub-band 1 | centered at 230 GHz, 128 channels equally spaced over the window [229 GHz, 231 GHz]; output: ALL 128 CHANNELS SIMULTANEOUSLY (middle 100-110 useable) |
| sub-band 2 | centered at 232 GHz, 128 channels equally spaced over the window [231 GHz, 233 GHz]; output: ALL 128 CHANNELS SIMULTANEOUSLY (middle 100-110 useable) |
| sub-band 3 | centered at 246 GHz, 128 channels equally spaced over the window [245 GHz, 247 GHz]; output: ALL 128 CHANNELS SIMULTANEOUSLY (middle 100-110 useable) |
| sub-band 4 | centered at 248 GHz, 128 channels equally spaced over the window[247 GHZ, 249 GHz; output: ALL 128 CHANNELS SIMULTANEOUSLY (middle 100-110 useable) |

***Table 8.1:*** *The two receiver bands and sub-bands likely to be offered during Cycle 4.*

| FOV | 6x6 SMALL | 9x9 MEDIUM | 12x12 LARGE |
|---|---|---|---|
| Band 3 | 2.8' x 2.8' (1.4' x 1.4') | 4.3' x 4.3' (2.1' x 2.1') | 5.7' x 5.7' (2.8' x 2.8') |
| Band 6 | 1.2' x 1.2' (0.6' x 0.6') | 1.8' x 1.8' (0.9' x 0.9') | 2.3' x 2.3' (1.2' x 1.2') |
| Time | 4.5 min + 3min cal. | 16.1 min (13 min*) | 24.0 min |

*Table 8.2: Field of view and time for completion for quadratic mosaics (examples only). The FOV sizes are for step sizes of 1/2 the beam size and (those in brackets) for 1/4 the beam size.*
*\* The medium 9x9 mosaic might fit exactly in a single 10min block.*

- The biggest mosaic would take ~26min (2 observing blocks of 10 min each plus 2 calibration breaks of 3 min each).
- There is a trade-off between FOV size and cadence for mosaicking, which has to be evaluated with respect to the requirements of each individual science case. Please see Fig. 2 and Table 8.2. for examples for mosaics with quadratic FOVs.

## 6. Sequence duration and calibration

An observation (i.e. a whole ALMA proposal) comprises of several (or one) **Science Goals**. A Science goal can last for up to 2hours (incl. 30min initial calibration) and is a set of **Scheduling Blocks** (SB), and SBs are separated into **Executing Blocks** (EB). An EB is a basic observation unit, which includes the calibration of the bandpass, sideband separation, flux, amplitude and phase and the actual science observation. A Science Goal is thus a group/sequence of observations that is designed for a particular science topic but by **using one band at a time only** for now.

It is currently assumed that solar science observations have a default **maximum duration** of 10min.

ALMA will target a **calibrator source** between two science observations, which will take ~3min (as of now).

A sequence that alternates between band 3 and 6 could be realized by defining band 3 and band 6 observations as different Science Goals and switching between these goals, although there would be substantial breaks for calibration. Possible sequences for the suggested standard observing set-ups (see Table 6.2) are illustrated in Table 8.3 on the next page.

*In addition, **in-band** fast switching for calibration purposes <u>might</u> be possible. That means that changing between the source and phase calibrator can be done quicker than 5min. This capability has been tested during the long baseline campaign, resulting in fast switching within 10-30sec. However, this feature may not greatly improve the calibration but reduces the time for actual science observations even further.*

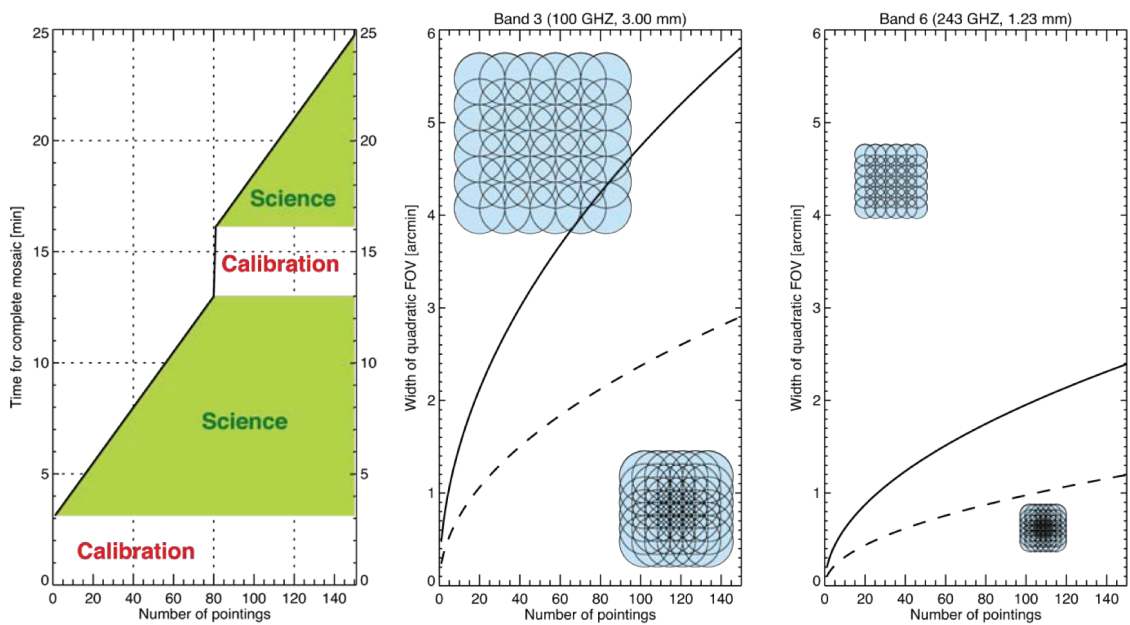

*Figure 2: Mosaicking in Cycle 4. The leftmost panel shows the time it takes to complete a mosaic with a given number of pointings. The width of the corresponding field of views (here assumed to be quadratic) is plotted in the middle panel for band 3 and in the rightmost panel for band 6. For both bands, mosaics with a step size of 1/2 (solid line, recommended Nyquist sampling) and 1/4 (dashed line) times the primary beam are shown, respectively. See Table 8.2 for examples. The durations for each science observing block and intermediate calibration block are assumed to be 10 min and 3 min, respectively, while the time per pointing is set to 7.5 s.*

|  | SD | INTERFEROMETRIC IMAGING | | | | | |
|---|---|---|---|---|---|---|---|
| ID | SD-1 | INT-1 | INT-2 | INT-3 | INT-4 | INT-5 | INT-6 |
| Name | FULL-DISK 1 | FAST | MODERATE | SMALL MOSAIC | LARGE MOSAIC | MULTI | SWEEP |
| 0:00 | Prep. | Array + receiver setup Band 6 | Array + receiver setup Band 6 | Array + receiver setup Band 6 | Array + receiver setup Band 6 | Array + receiver setup Band 6 | Array + receiver setup Band 6 |
| 0:05 | Initial calibration | Initial calibration | Initial calibration | Initial calibration | Initial calibration | Initial calibration | Initial calibration |
| 0:30 | Calibration | Calibration | Calibration | Calibration | Calibration | Calibration | Calibration |
| 0:33 | Full disk scan #1 Band 6 1 map | Science Band 6 cadence 2 s 300 maps | Science Band 6 cadence 10 s 60 maps | Science Band 6 2 mosaics 6x6 pointings | Science Band 6 1. mosaic pointing 1-80 | Science Band 6 pointing #1 cadence 2 -10s 60-300 maps | Science Band 6 pointing 1-10 (1 min each) cadence 2 -60s 60 - 300 maps |
| 0:43 | Calibration | Calibration | Calibration | Calibration | Calibration | Calibration | Calibration |
| 0:46 | Full disk scan #2 Band 6 1 map | Science Band 6 cadence 2 s 300 maps | Science Band 6 cadence 10 s 60 maps | Science Band 6 2 mosaics 6x6 pointings | Science Band 6 1. mosaic pointing 81-150 | Science Band 6 pointing #2 cadence 2 -10s 60-300 maps | Science Band 6 pointing 11-20 (1 min each) cadence 2 -60s 60 - 300 maps |
| 0:56 | Calibration | Calibration | Calibration | Calibration | Calibration | Calibration | Calibration |
| 0:59 | Full disk scan #3 Band 6 1 map | Science Band 6 cadence 2 s 300 maps | Science Band 6 cadence 10 s 60 maps | Science Band 6 2 mosaics 6x6 pointings | Science Band 6 2. mosaic pointing 1-80 | Science Band 6 pointing #3 cadence 2 -10s 60-300 maps | Science Band 6 pointing 21-30 (1 min each) cadence 2 -10s 60 - 300 maps |
| 1:09 | Calibration | Calibration | Calibration | Calibration | Calibration | Calibration | Calibration |
| 1:12 | Full disk scan #4 Band 6 1 map | Science Band 6 cadence 2 s 300 maps | Science Band 6 cadence 10 s 60 maps | Science Band 6 2 mosaics 6x6 pointings | Science Band 6 2. mosaic pointing 81-150 | Science Band 6 pointing #4 cadence 2 -10s 60-300 maps | Science Band 6 pointing 31-40 (1 min each) cadence 2 -10s 60 - 300 maps |
| 1:22 | Calibration | Calibration | Calibration | Calibration | Calibration | Calibration | Calibration |
| 1:25 | Full disk scan #5 Band 6 1 map | Science Band 6 cadence 2 s 300 maps | Science Band 6 cadence 10 s 60 maps | Science Band 6 2 mosaics 6x6 pointings | Science Band 6 3. mosaic pointing 1-80 | Science Band 6 pointing #5 cadence 2 -10s 60-300 maps | Science Band 6 pointing 41-50 (1 min each) cadence 2 -10s 60 - 300 maps |
| 1:35 | Calibration | Calibration | Calibration | Calibration | Calibration | Calibration | Calibration |
| 1:38 | Full disk scan #6 Band 6 1 map | Science Band 6 cadence 2 s 300 maps | Science Band 6 cadence 10 s 60 maps | Science Band 6 2 mosaics 6x6 pointings | Science Band 6 3. mosaic pointing 81-150 | Science Band 6 pointing #6 cadence 2 -10s 60-300 maps | Science Band 6 pointing 51-60 (1 min each) cadence 2 -10s 60 - 300 maps |
| 1:48 | Calibration | Calibration | Calibration | Calibration | Calibration | Calibration | Calibration |
| 1:51 | *Enough time to complete a Band 6 scan?* | Science Band 6 cadence 2 s 210 maps | Science Band 6 cadence 10 s 42 maps | Science Band 6 1 mosaic=5min 6x6 pointings | *Not enough time for mosaic left* | Science Band 6 pointing #7 cadence 2 -10s 54-270 maps | Science Band 6 pointing 61-67 (1 min each) cadence 2 -10s 54 - 270 maps |
| 2:00 | Next Science Goal: Switching to receiver band 6 (if requested) and executing the same sequence for band 3 | | | | | | |

*Table 8.3:* Possible Science Goals (complete observing sequences) for suggested solar observing set-ups in Cycle 4. This example shows sequences in Band 6 but could be done in the same way for Band 3 (Note that SD scans in Band 6 take only 7min).

The max. duration of a Science Goal (incl. calibration) is set to 2 hours. Array and receiver setup are assumed to take less than 5 min. The "initial calibration" (bandpass, sideband separation, flux, pointing) is assumed to take 25min. The frequent (intermediate) calibration includes phase, atmosphere ($T_{sys}$), and amplitude calibration and is assumed here to take 3 min.

**Please note that the timings in these examples are only tentative.**

ALMA OBSERVATIONS OF THE SUN IN CYCLE 4 AND BEYOND## 9. CO-OBSERVING WITH OTHER OBSERVATORIES

Co-observing has become a standard procedure in contemporary solar physics. It is scientifically motivated by the insight that the different layers of the solar atmosphere are intricately coupled so that simultaneous observations of as many layers as possible are needed in order to enable true understanding of the physical processes at work. For this purpose, different observatories on the ground and in space are usually co-observing the same target region on the Sun, providing complementary plasma diagnostics and/or context data. ALMA is not just yet another observatory joining multi-instrument campaigns but rather provides new diagnostic tools of central importance for the chromosphere, sometimes also called the "interface region", and thus for a large range of scientific topics. In return, complementary observations of different layers of the Sun with other instruments amplify the scientific potential of ALMA for the Sun. Consequently, it can be expected that every solar ALMA campaign would be followed by a number of other instruments covering a large range of the spectrum.

In the following, we give a short overview of currently available instruments, although this list is not exhaustive. Many smaller solar telescopes, which are not mentioned, might be able to join ALMA campaigns, too. Please also refer to the list given in the review article mentioned in Sect. 1.

**Space-borne solar observatories**

The **Interface Region Imaging Spectrograph (IRIS)** is has committed to supporting ALMA at the highest priority during the months of December and January and will also support ALMA in February and March at high priority, if coordination with other observatories and telemetry allow. IRIS obtains images and spectra in the near-UV and far-UV at high spatial (0.33"), temporal (~2s) and spectral resolution (~3 km/s). Many of the spectral lines observed by IRIS (e.g., Mg II h and k) and continua are formed in the same region as the millimeter continua observed with ALMA, but probe different plasma properties, thereby providing a highly complementary dataset to the ALMA data. Because of the daily planning cycle of IRIS, simultaneous observations of ALMA and IRIS require target and observing program coordination one day before the actual observing. Co-observing of ALMA and IRIS has already been successfully demonstrated during the *Commissioning and Science Verification* campaign in December 2014.

The **Hinode** spacecraft can provide complementary data for the photosphere below (incl. photospheric magnetograms) and the transition region and corona above the region probed by ALMA. Hinode's FOV is much bigger than ALMA's FOV but still requires exact coordination of the target coordinates like it is the case for IRIS. During the months of February-April and August-October, IRIS and Hinode are committed to close coordination through so-called IRIS Hinode Operation Plans (IHOPs). Especially during this spring time period, inclusion of ALMA in these coordination plans would serve the interests of all three projects. *(Outside these periods, Hinode time must probably be applied for separately and well in advance.)*

The **Solar Dynamics Observatory (SDO)** is continuously observing the full disk of the Sun at all times in many wavelengths bands at a cadence of down to 12s. It thus provides context information about the corona above the layers probed by ALMA and the photosphere below (incl. photospheric magnetograms). SDO data is publically available and can always be obtained after an ALMA observation without the need for further coordination - in contrast to Hinode and IRIS. SDO observations can directly be compared with ALMA's TP full disk scans

The primary mission of the **Reuven Ramaty High Energy Solar Spectroscopic Imager (RHESSI)** is the study of solar flares by using imaging spectroscopy in the X-ray and gamma-ray range. RHESSI has a high angular and energy resolution, which provides crucial information on particles accelerated during flares and the resulting non-thermal radiation. ALMA would complement RHESSI observations not only with diagnostics for the thermal component of radiation emitted during flares but also provides important constraints on the non-thermal component. RHESSI observes the whole disk of the Sun at all times so that no explicit coordination of the two instruments is necessary. Rather, RHESSI data is available whenever ALMA captures a solar flare.

http://ssalmon.uio.no     21

### Ground-based observatories

The **New Solar Telescope** (NST) at Big Bear Solar Observatory in California, USA, has an aperture of 1.6 m, which is currently the world's largest, followed by the **GREGOR** telescope on Tenerife, La Palma, with an aperture of 1.5 m. The **Swedish 1-m Solar Telescope (SST)** is also located on the Canary Islands (La Palma), has 1-m aperture but post-focus instruments (CRISP and in future CHROMIS), which can be considered some of the foremost workhorses in chromosphere studies. The **Vacuum Tower Telescope** (**VTT**, 0.7 m aperture) and **THEMIS** (0.9 m aperture) are located on Tenerife, too. The **Dunn Solar Telescope (DST)** in New Mexico, USA, is still in use and offers instruments such as IBIS and ROSA, which are interesting for studying the chromosphere. All these optical telescopes can access wavelengths from about 400 nm up to typically 1μm (or even 12 μm in the case of GREGOR) and offer a large variety of diagnostic tools from high-cadence imaging to spectroscopy.

The **H-alpha Solar Telescope for Argentina** (**HASTA**) is located in San Juan (Argentina) in the Andes not too far from ALMA and thus in the same time zone. HASTA could provide complementary Hα observations of the solar chromosphere. The spatial resolution is 1.5 arcsec and the temporal resolution is 3 seconds. HASTA works in "patrol mode".

The **Solar Submillimeter-wave Telescope** (SST, not to be confused with the Swedish 1-m Solar Telescope) is located at the same site as HASTA and can observe the Sun at frequencies of 212 GHz and 405 GHz, which coincide with ALMA bands 6 and 8, respectively.

### Next generation optical solar telescopes

The **D.K. Inouye Solar Telescope (DKIST)** with its 4 m aperture is expected to see first light in 2019 (i.e., during ALMA Cycle 6-7). Its location on Hawaii would limit simultaneous observing with ALMA to 1-2 hours, which is still a usual duration for most high-impact science cases. The future European Solar Telescope (EST) would most likely be located on the Canary Islands and would have an aperture of 4 m, too.

### Conditions for co-observing

While space-borne telescopes are in principle available 24/7 unless in eclipse season, coordination with ground-based instruments is subject to weather conditions. Furthermore, the time difference due to the different locations has to be taken into account, at least when simultaneous co-observing is required. During the most likely solar ALMA observing season (southern summer, see Sect. 8.1) ALMA is in principle available from 13:00 to 20:00UT. The time difference to the telescopes on the Canary Islands (e.g., GREGOR, SST, VTT, THEMIS) is about 4 hours ahead. Best seeing conditions are usually in the morning, thus before ALMA observations can begin. The NST and the DST are about 3-4 hours behind ALMA time and could thus co-observe in the morning-early afternoon (NST time). For all aforementioned telescopes a few hours of simultaneous observing with ALMA should be possible.

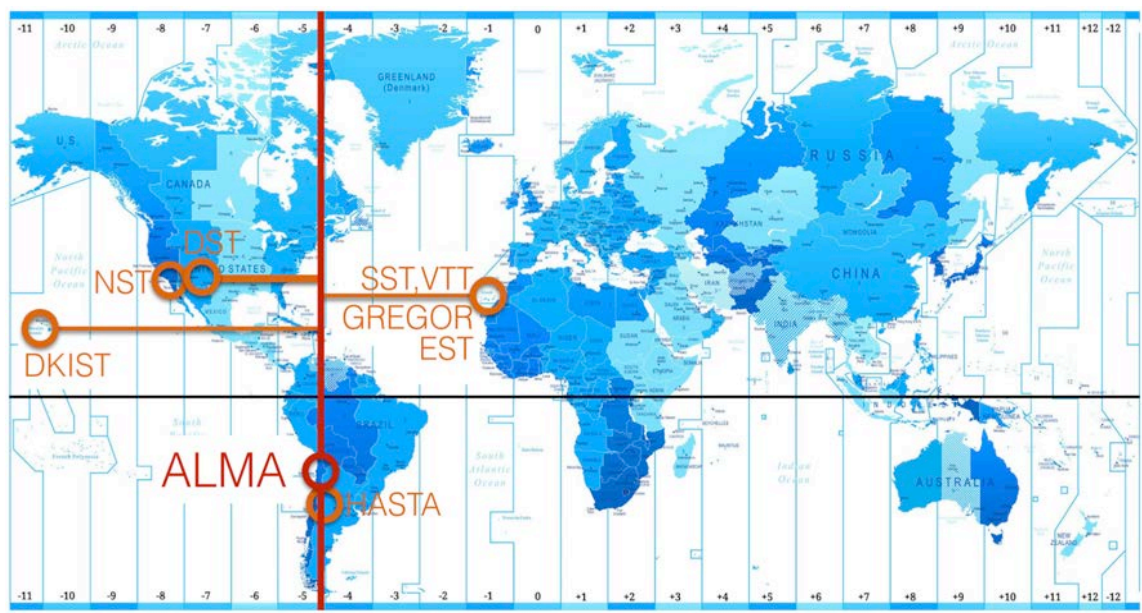

***Figure 3:*** *Geographic location of ALMA and some of current and future ground-based optical solar telescopes, which are anticipated to participate in coordinated observing campaigns.*

## 10. QUIET SUN REGIONS (TEAM K)

**List of contributing team members:**
T. Ayres, M. Carlsson, G. Harper, S. Jafarzadeh, M. Loukitcheva, S. Wedemeyer

**Lead author:** S. Wedemeyer *(sven.wedemeyer@astro.uio.no)*

### K1 - SUMMARY OF SCIENCE CASES AND KEY QUESTIONS

**Science Case K-SC1: The brightness temperature distribution in quiet Sun regions and its center-to-limb variation**

- **Key Question SC-K1-KQ1:** What are the statistical characteristics of the brightness temperature distribution in quiet Sun regions?
- **Key Question SC-K1-KQ2:** How does the brightness temperature distribution depend on the observation angle (and thus radial distance from the solar disk center)?
- **Key Question SC-K1-KQ3:** How do the measured brightness temperatures depend on the integration time of the observations?

**Science Case K-SC2: Local slopes of the brightness temperature spectrum**

- **Key Question SC-K2-KQ1:** What is the local slope of the brightness temperature spectrum as function of observing frequency and thus sampled atmospheric height?

**Science Case K-SC3: Turbulent diffusion and velocity fields above supergranules**

- **Key Question SC-K3-KQ1:** What are the characteristics of horizontal flows within a supergranulation cell and how do they vary with height?
- **Key Question SC-K3-KQ2:** How is the velocity spectrum in the quiet-Sun chromosphere characterized?

### K2 - DETAILED DESCRIPTIONS OF SCIENCE CASE

#### K2.1 - Science Case SC-K1: The brightness temperature distribution in quiet Sun regions and its center-to-limb variation

**SC-K1-KQ1: What are the statistical characteristics of the brightness temperature distribution in quiet Sun regions?**

As an essential first step, the brightness temperatures in quiet Sun regions should be measured systematically. The measurements would rely on reconstructed interferometric ALMA maps with the highest spatial resolution achievable at the time. Each map would result in a brightness temperature histogram and basic properties such as dynamic range, mean value and standard deviation (and with it the contrast). Given the rather small FOV, observations over 5-10 min per pointing will increase the statistical significance while at the same time give a first impression of the atmospheric dynamics as they appear in the ALMA bands.

Comparisons of these results to numerical models of the solar chromosphere will allow us to investigate several important points:

(i) Can the available models already realistically reproduce the mm observations or are there discrepancies, which need to be investigated further?

(ii) Is the spatial resolution of the ALMA maps already sufficient to capture the details of the brightness temperature distribution on small spatial scales?

Band 3 (~100GHz, 3mm) and band 6 (~240GHz, 1.2mm) probe mm continuum radiation from slightly different heights in the solar chromosphere. Based on numerical 3D models, band 6 would probe a layer at an average height of ~700k m whereas band 3 would probe a layer at ~900 km, depending on the employed model. Detailed models with time-dependent treatment of hydrogen ionisation imply average formation heights of ~750 km and ~960 km for bands 6 and 3, respectively. However, the height ranges would certainly overlap for both bands. Co-observations with other instruments, here in particular IRIS, and detailed comparisons of the structures in the observed maps could help to further constrain the actual formation height ranges of ALMA bands 3 and 6.

The proposed ALMA observations would serve as a crucial test for the models. *(The required detailed comparison should be carried out in collaboration with teams A, B, C).* See also science case SC-K2. Furthermore, these observations would also allow for analysing the accuracy of the brightness temperature calibration.

**SC-K1-KQ2: How does the brightness temperature distribution depend on the observation angle (and thus radial distance from the solar disk center)?**
Observations at several positions from disk-center to the limb add more constraints on the probed thermal structure of the chromosphere and should be used as a crucial test for all numerical and semi-empirical models of the Sun's chromospheric stratification. The inclination angle ($\mu = \cos\theta$) of the line-of-sight changes from $\theta=0°$ ($\mu=1.0$, vertical) at disk-center to $\theta=90°$ ($\mu=0.0$, sideways) at the limb. Consequently, the height at which the chromosphere becomes optically thick at a given mm wavelength (and thus the layer from where most of the mm radiation is emitted) increases from disk-center to the limb. This effect is seen in numerical models. Detailed comparisons would therefore provide another important test.

**SC-K1-KQ3: How do the measured brightness temperatures depend on the integration time of the observations?**
Although the required integration time should be low, we intend to remain 5min on each position. After the observations, maps can be reconstructed by using all information from time windows with different durations, thus in the extreme case using information from the total duration of 5min. Comparing the resulting maps with these different overall "integration times" would reveal the following two important points:

(i) What is the longest time window that can be used for the image reconstruction without losing (i.e., smearing out) details of the dynamic small-scale pattern (e.g., propagating shock waves)? Simulations imply that the relevant timescales are on the order of a few 10 s only.

(ii) How is the brightness temperature distribution affected by too long integration times and how could this be corrected for?

**Observational requirements**

- **Target:** FOVs with quiet Sun regions at positions from disk-center to the limb.
- **Array configuration:** Interferometric observing in compact configuration plus continuous fast scanning with 3-4 TP antennas (FOV 2400" x 2400").

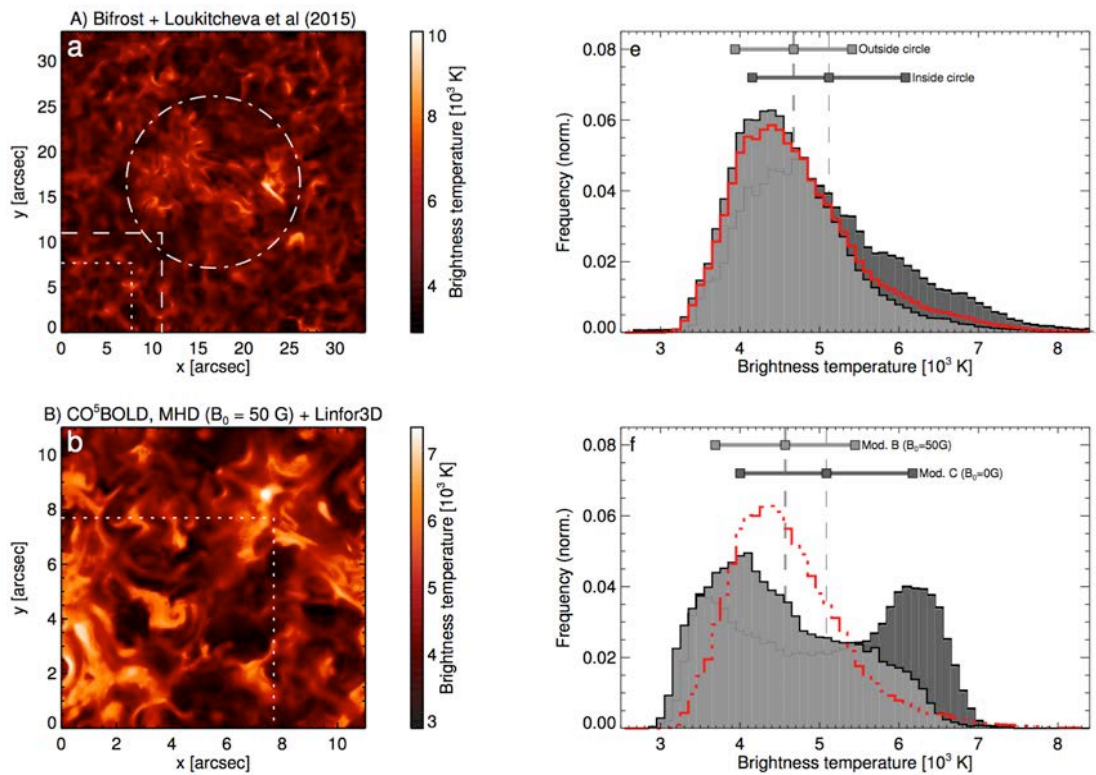

***Figure K.1:*** *(Left) Synthetic brightness temperature maps at a wavelength of 1mm (band 6) based on two different numerical 3D models of the solar chromosphere. (Right) Histograms of the synthetic brightness temperatures. From Wedemeyer et al. (2015, SSRv).*

- **Pointing/FOV:** Sequence of 10 single pointings starting at disk center (see Table K1).
- **Receiver:** The observation sequence should be carried out first with band 3 and then with band 6. Full advantage will be made of the sub-bands with 100-110 useable spectral channels in interferometric mode and the additional 4 sub-bands from the simultaneous TP observing. *(Backup: The study could be carried out with only one frequency per band but using the 4 simultaneous sub-bands is preferable, in particular for science case SC-K2, see Sect. 2.2).*
- **Temporal resolution:** The observations should have a cadence of less than 10s in order to avoid smearing out of the chromospheric small structure. Ideally the highest cadence (2sec) should be used but only if the signal is strong enough in the darker parts of the targeted Quiet Sun regions.
- **Duration:** Completing the pointing sequence for each band would take 130min incl. all calibrations.
- **Co-observing with IRIS:** Highly preferable. Due to the pointing sequence (10 per band) a strict schedule should be followed.

**Proposed observation sequence**

The observations would begin at disk-center and proceed towards the limb in 10 steps with equal increment in view inclination angle ($\Delta\mu = 0.1$). Both available bands (3+6) should be used at each position. Due to large overheads when changing receiver bands, we propose to carry out the center to limb steps first in band 3 and then band 6. Both are defined as separate Science Goals (SG). In order to keep the overall duration short and thus increasing chances for being selected, each SG is executed only once. At each position, we intend to observe for 5min and with the shortest cadence possible (1-2) sec. TP fast scanning will continue in the same band for the full duration of the interferometric observations.

**Technical considerations**

- Is the S/N ratio for observation in the darkest parts of Quiet Sun regions good enough at a "integration time" of 1-2sec or should it be increased to up to 10sec? In any case, each position should be observed for 5min in order to allow for experiments with the effective "integration time" afterwards
- The observation sequences for each band take 65 min (incl. 15 min calibration see Table K1).

- **Please also see the following science cases, which target Quiet Sun regions:**
- **SC-G1:** Acoustic waves
- **SC-J2:** Dynamic nature of atmospheric heating
- **SC-K1:** The brightness temperature distribution in quiet Sun regions and its center-to-limb variation
- **SC-K2:** Local slopes of the brightness temperature spectrum
- **SC-K3:** Turbulent diffusion and velocity fields above supergranules
- **SC-P1:** The mean vertical structure of the solar atmosphere
- **SC-P2:** Variation of the brightness temperature on network and granule scales

**K2.2 - Science Case SC-K2: Local slopes of the brightness temperature spectrum**

**SC-K2-KQ1: What is the local slope of the brightness temperature spectrum as function of observing frequency and thus sampled atmospheric height?**

The same observations as proposed for SC-K1 could be used to investigate the dependence of the brightness temperatures on the observing frequency. As already mention for SC-K1-KQ1 in Sect. 2.1, the formation height of the observed mm radiation should on average increase with wavelength (decrease with observing frequency). The difference in formation height between band 3 and 6 is expected to be on average on the order of 200km - 500km, depending on the employed model. Comparing the proposed ALMA observations with models will help to better constrain the different formation height ranges.

For band 3, sub-bands at the following frequencies will be available: 94, 96, 104 and 106 GHz, corresponding to the wavelengths 3.19 mm, 3.12 mm, 2.88 mm, 2.83 mm, respectively. The difference between the provided sub-bands will therefore be even smaller and certainly smaller than the formation height differences between different positions in the FOV. Nevertheless, the brightness temperatures measured simultaneously in the 4 sub-bands allow to determine the slope of the continuum, which carries information on the thermal structure along the line of sight. For both, the interferometric observations and the TP fast scans, a total of 2 x 4 data points (and even 100-110 spectral channels for the sub-bands in interferometric observing) would be available for attempting a rough reconstruction of the height profile of the gas temperature in the probed chromospheric layers for all positions across the FOV.

The proposed ALMA observations would serve as a crucial test for numerical models. *(The required*

*detailed comparison should be carried out in collaboration with teams A, B, C).*

**Observational requirements**

Same as for SC-K1 (see Sect. K2.1).

- **Please also see the following science cases, which target Quiet Sun regions:**
- **SC-G1:** Acoustic waves
- **SC-J2:** Dynamic nature of atmospheric heating
- **SC-K1:** The brightness temperature distribution in quiet Sun regions and its center-to-limb variation
- **SC-K2:** Local slopes of the brightness temperature spectrum
- **SC-K3:** Turbulent diffusion and velocity fields above supergranules
- **SC-P1:** The mean vertical structure of the solar atmosphere
- **SC-P2:** Variation of the brightness temperature on network and granule scales

### K2.3 - Science Case SC-K3: Turbulent diffusion and velocity fields above supergranules

**SC-K3-KQ1: What are the characteristics of horizontal flows within a supergranulation cell and how do they vary with height?**

Recent high-resolution observations of the solar photosphere with various modern telescopes (both on the ground and in space) have advanced our understanding of turbulent diffusion in supergranulation cells by utilizing magnetic features as tracers of horizontal flow field, which is advecting the features. These measurements have been lacking information from higher atmospheric layers above the supergranules where the flow field is believed to be guided by prevailing (horizontal) magnetic fields. Thus, a chromospheric "diffusion zone" could be formed as a result of plasma stabilisation by the horizontal fields. Tracking bright (magnetic) features and using advanced local-correlation-tracking (LCT) algorithms applied to chromospheric observations with ALMA (in both Band 3 and Band 6) would enable us to analyse the diffusion in these layers. As a consequence, we will be able to improve our understanding of, e.g., mass and energy flows in the chromosphere, as well as their relationships with other atmospheric layers. The anticipated results have therefore implications for our understanding of the energy budget of the upper solar atmosphere.

Co-observations of photospheric/chromospheric layers with, e.g., Hinode and/or SST, as well as transition regions with IRIS would enable us to more precisely probe the nature of the turbulence as well as the relationship between kinetic energy density of the horizontal flows (within the atmosphere) and the magnetic energy density.

**Key Question SC-K3-KQ2: How is the velocity spectrum in the quiet-Sun chromosphere characterized?**

Velocity spectra, in particular spectra for horizontal velocities, have been computed for different spatial scales in the solar photosphere based on observations with, e.g., SOHO, SDO, Hinode, and Sunrise. ALMA observations can provide similar data but for many more simultaneous heights in the atmosphere since the probed height increases with the observed wavelength. An LCT analysis of these data would therefore produce the velocity spectrum as function of height for the sampled chromospheric layers. These measured spectra at various atmospheric heights would result in a better view of velocity fields in the solar atmosphere.

**Observational requirements**

Similar to that of SC-K1 and SC-K2 (see Sects. K2.1-K2.2), but with high cadence and the longest possible time-series of images (both band 3 and band 6). The length of observations is an important factor in this project although it is feasible with also relatively short image-sequences at the expense of statistics and accuracy.

- **Please also see the following science cases, which target Quiet Sun regions:**
- **SC-G1:** Acoustic waves
- **SC-J2:** Dynamic nature of atmospheric heating
- **SC-K1:** The brightness temperature distribution in quiet Sun regions and its center-to-limb variation
- **SC-K2:** Local slopes of the brightness temperature spectrum
- **SC-K3:** Turbulent diffusion and velocity fields above supergranules
- **SC-P1:** The mean vertical structure of the solar atmosphere
- **SC-P2:** Variation of the brightness temperature on network and granule scales

| | | | Duration | Time |
|---|---|---|---|---|
| | **Start of EB. Science Goal A (BAND 3)** | | | |
| | A0s | Make Array and receiver setup for **Band 3** | 5min | 0:00 |
| | A0c | Bandpass, sideband separation, flux, pointing calibration for Band 3 | 25min | 0:05 |
| | A1TPc | Total Power antenna calibration for band 3 | 3min | 0:30 |
| | A1c | Phase, atmosph. ($T_{sys}$), amp. calibrat. for Band 3; move to target (μ=1.0) | 3min | 0:30 |
| | A1TPo | **Total Power science observation in band 3**: continuous fast scanning, | (62min) | 0:33 |
| | A1o1 | **Science observation** for Band 3 - Target: disk center (μ=1.0) | **5min** | 0:33 |
| | A1o2 | **Science observation** for Band 3 - Target: μ=0.9 | **5min** | 0:38 |
| | A2c | Phase, atmosph. ($T_{sys}$), amp. calibrat. for Band 3; move to target (μ=0.8) | 3min | 0:43 |
| | A2o1 | **Science observation** for Band 3 - Target: μ=0.8 | **5min** | 0:46 |
| | A2o2 | **Science observation** for Band 3 - Target: μ=0.7 | **5min** | 0:51 |
| | A3c | Phase, atmosph. ($T_{sys}$), amp. calibrat. for Band 3; move to target (μ=0.6) | 3min | 0:56 |
| | A3o1 | **Science observation** for Band 3 - Target: μ=0.6 | **5min** | 0:59 |
| | A3o2 | **Science observation** for Band 3 - Target: μ=0.5 | **5min** | 1:04 |
| | A4c | Phase, atmosph. ($T_{sys}$), amp. calibrat. for Band 3; move to target (μ=0.4) | 3min | 1:09 |
| | A4o1 | **Science observation** for Band 3 - Target: μ=0.4 | **5min** | 1:12 |
| | A4o2 | **Science observation** for Band 3 - Target: μ=0.3 | **5min** | 1:17 |
| | A5c | Phase, atmosph. ($T_{sys}$), amp. calibrat. for Band 3; move to target (μ=0.4) | 3min | 1:22 |
| | A5o1 | **Science observation** for Band 3 - Target: μ=0.2 | **5min** | 1:25 |
| | A5o2 | **Science observation** for Band 3 - Target: μ=0.1 | **5min** | 1:30 |
| | **End of the EB and Science Goal A** | | | 1:35 |
| | | | | |
| | **Start of EB. Science Goal B (BAND 6)** | | | 1:35 |
| | B0s | **Switch to receiver band 6**; make Array and receiver setup for **Band 6** | 5min | 1:35 |
| | B0c | Bandpass, sideband separation, flux, pointing calibration for Band 6 | 25min | 1:40 |
| | B1TPc | Total Power antenna calibration for band 6 | 3min | 2:05 |
| | B1c | Phase, atmosph. ($T_{sys}$), amp. calibrat. for Band 6; move to target (μ=1.0) | 3min | 2:05 |
| | B1TPo | **Total Power science observation in band 6**: continuous fast scanning, | (62min) | 2:08 |
| | B1o1 | **Science observation** for Band 6 - Target: disk center (μ=1.0) | **5min** | 2:08 |
| | B1o2 | **Science observation** for Band 6 - Target: μ=0.9 | **5min** | 2:13 |
| | B2c | Phase, atmosph. ($T_{sys}$), amp. calibrat. for Band 6; move to target (μ=0.8) | 3min | 2:18 |
| | B2o1 | **Science observation** for Band 6 - Target: μ=0.8 | **5min** | 2:21 |
| | B2o2 | **Science observation** for Band 6 - Target: μ=0.7 | **5min** | 2:26 |
| | B3c | Phase, atmosph. ($T_{sys}$), amp. calibrat. for Band 6; move to target (μ=0.6) | 3min | 2:31 |
| | B3o1 | **Science observation** for Band 6 - Target: μ=0.6 | **5min** | 2:34 |
| | B3o2 | **Science observation** for Band 6 - Target: μ=0.5 | **5min** | 2:39 |
| | B4c | Phase, atmosph. ($T_{sys}$), amp. calibrat. for Band 6; move to target (μ=0.4) | 3min | 2:44 |
| | B4o1 | **Science observation** for Band 6 - Target: μ=0.4 | **5min** | 2:47 |
| | B4o2 | **Science observation** for Band 6 - Target: μ=0.3 | **5min** | 2:52 |
| | B5c | Phase, atmosph. ($T_{sys}$), amp. calibrat. for Band 6; move to target (μ=0.4) | 3min | 2:57 |
| | B5o1 | **Science observation** for Band 6 - Target: μ=0.2 | **5min** | 3:00 |
| | B5o2 | **Science observation** for Band 6 - Target: μ=0.1 | **5min** | 3:05 |
| | **End of the EB and Science Goal B** | | | 3:10 |

*Table K1: Proposed observing sequence for science case K-SC1 with 5min blocks at positions from disk-center to the limb first in band 3 and then in band 6. The observations in the two bands are defined as two individual Science Goals.*

## 11. ACTIVE REGIONS (TEAM L)

**List of contributing team members:**
G. Fleishman, M. Loukitcheva, S. Wedemeyer

**Lead author**: *G. Fleishman  (gfleishm@njit.edu)*

### L1 - SUMMARY OF SCIENCE CASES AND KEY QUESTIONS

Our Sun is active and its activity is variable in time. The most well-known period of the activity is the 11-years cycle, which can be quantified by various indices of solar activity; one of which is the so-called Wolf number, defined by the number of sunspots and groups of sunspots. A sunspot is relatively big area on the solar surface (i.e., the photosphere), which appears darker than the surrounding 'quiet' sun. The darkest area in the center of a sunspot is called 'umbra,' while the less dark, filamentary belt around it is called 'penumbra'. It is known that the darkening of sunspots is due to strongly enhanced magnetic fields, which inhibit convection and, thus, the heat transfer from below, resulting in a lower temperature of the sunspot than the surrounding quiet sun. An area covering a group of sunspots and/or other 'magnetically active' elements of the photosphere is called 'Active Region'. In Active Regions magnetic energy is highly enhanced, compared to the surroundings, and is responsible for most of the solar activity phenomena including solar flares and coronal mass ejections.

**Science Case SC-L1: The brightness temperature distribution in active regions and sunspots**

- **Key Question SC-L1-KQ1:** What are the statistical characteristics of the brightness temperature distribution in flare-productive and non-flaring active regions?

**Science Case SC-L2: The dependence of Active Region brightness temperatures on magnetic field strength**

- **Key Question SC-L2-KQ1:** How does the brightness temperature spectrum depend on the magnetic field properties of the underlying photospheric component (umbra, penumbra, plage, facula, etc.)?
- **Key Question SC-L2-KQ2:** How does it agree with static and dynamic models of the solar atmosphere?

**Science Case SC-L3: Height-dependence of the brightness temperature spectrum in Active Regions**

- **Key Question SC-L3-KQ1:** What is the local slope of the brightness temperature spectrum as function of observing frequency and thus sampled atmospheric height?
- **Key Question SC-L3-KQ2:** What is the height scale of Active Region structure changes?

**Science Case SC-L4: Temporal evolution of Active Regions**

**Key Question SC-L4-KQ1:** How does the brightness temperature spectrum varies with time for a given atmospheric component (umbra, penumbra, plage, facula, etc.)?

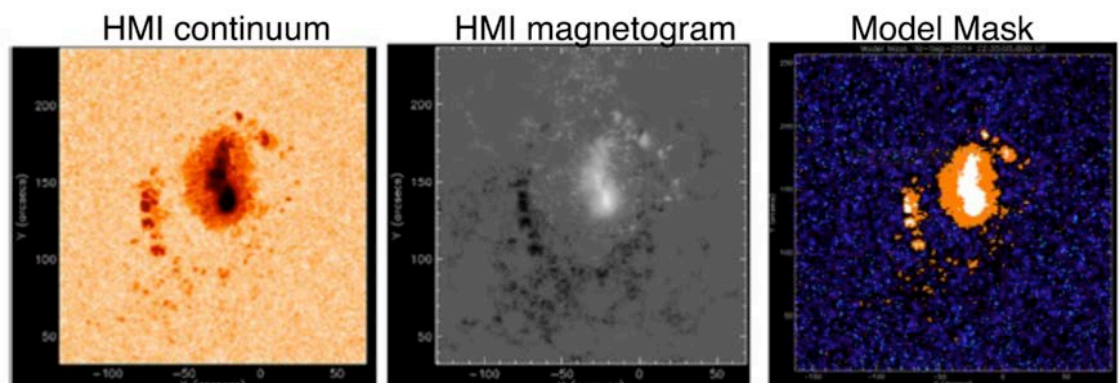

***Figure L.1:*** *Images of AR 12158 from HMI/SDO on Sep 10, 2014. From left to right: white light image, photospheric magnetogram and model mask made based on the photospheric input.*

## L2 - DETAILED DESCRIPTIONS OF SCIENCE CASES

### L2.1 - Science Case SC-L1: The brightness temperature distribution in active regions and sunspots

**Key Question SC-L1-KQ1:** What are the statistical characteristics of the brightness temperature distribution in flare-productive and non-flaring active regions?

It is known that some active regions are highly flare-productive, while others are relatively quiet. On a general level it is established that more complex and dynamic active regions contain more free magnetic energy and so have more potential of producing a flare than magnetically "simple" regions. It is unknown, however, what are the statistical properties of these two general types of active regions at the chromospheric layers and to what extent their thermal structures are different.

**Observational requirements**

- **Summary:** Two regions of different activity types should be pre-selected based on the solar activity indexes for the observations with the same array setup.
- **Targets:** (1) Non-flaring Active region close to solar disk center. (2) Flare-productive Active region close to solar disk center

- **Array configuration:** Compact (C40-1-C40-3).
- **Pointing/FOV:** Mosaic with 6x6 pointings and a FOV of 3arcmin x 3 arcmin (Band 3) or 1.3 arcm x 1.3 arcmin (Band 6).
- **Receiver:** The observation sequence should be carried out first with band 3, then with band 6.
- **Temporal resolution:** 8-10min (the time it takes to complete the mosaic)
- **Duration:** 1 hours per band and per target (4 hours in total).

**Context observations**

- H-alpha: Images from HASTA (see Sect. 9) Images and real-time automatic flare detection from Kanzelhohe Observatory (Poetzi et al. 2015) located in Austria, pixel resolution is 1 arcsec and temporal resolution is 6 seconds (patrol mode).
- NST IR and optical data.
- Magnetograms and white light data from SDO

### L2.2 - Science Case SC-L2: The dependence of Active Region brightness temperatures on magnetic field strength

**Key Question SC-L2-KQ1:** How does the brightness temperature spectrum depend on the magnetic field properties of the underlying photospheric component (umbra, penumbra, plage, facula, etc.)?

**Key Question SC-L2-KQ2:** How does it agree with static and dynamic models of the solar atmosphere?

To get an idea of how a given Active Region looks at the ALMA wavelengths, Fleishman et al. (2015) explored the recently updated set of 1D models (1D distributions of the electron temperature and density along with non-LTE ionized and neutral hydrogen densities with height) of the solar atmosphere (Fontenla et al. 2009, 2014), distinguishing between umbra, penumbra, network, internetwork, enhanced network, facula, or plage. To apply a given static atmospheric model to a given line of sight, the algorithm analyzes the photospheric input (white light and magnetogram) to classify the pixel as belonging to one of the itemized above photospheric features and creates a corresponding model mask. Then, a 1D chromospheric model is added on top of each pixel, which forms a simplified 3D chromospheric model of the AR embedded in an extrapolated 3D magnetic data cube. Although simplified, these models are comprehensive enough to perform many tests; in particular, to compute anticipated averaged mm / sub-mm emission maps from a given AR as would be observed by ALMA. Some results of this modeling for AR 12158 (white light and magnetic data are taken on 10 Sep 2014) are shown in this section for ALMA bands 3 and 6.

This modeling suggests that photospheric features will be distinguishable at the ALMA frequencies with the brightness temperature Tb contrast up to a few hundred K. In line with the results of Loukitcheva et al. (2014), the umbra, which is dark (cool) at the high frequencies, appears bright (hot) at the lower frequencies.

To test the validity of this modeling that relies on our current knowledge of the chromospheric structure we need (i) to map the AR repeatedly to obtain time-averaged images as test for stationary atmospheric models (both SD and INT modes are needed) and (ii) to study the spectral shape of the emission along a given LOS to test the models at different frequencies and thus at different heights.

**Observational requirements**

- **Summary:**
  **Step 1.** Image a pre-selected AR @ four frequencies per band first in band 3 (~100 GHz) and then in band 6 (230 GHz) with ALMA.

**Step 2.** Obtain spectra along a number of representative lines of sight (corresponding to various underlying photospheric components, selected based on the context observations) taken at various times.

**Step 3.** Investigate the morphology of the AR, in comparison with context observations in other wavelengths and magnetograms to distinguish between different components the of thermal structure of the solar chromosphere.

- **Target:** Active region close to solar disk center.
- **Array configuration:** Compact (C40-1-C40-3).
- **Pointing/FOV:** Mosaic with 6x6 pointings and a FOV of 3arcmin x 3 arcmin (Band 3) or 1.3 arcm x 1.3 arcmin (Band 6).
- **Receiver:** The observation sequence should be carried out first with band 3 and then with band 6.
- **Temporal resolution:** 8-10min (the time it takes to complete the mosaic)
- **Duration:** 1-2 hours per band.

**Context observations**

- H-alpha: Images from HASTA (see Sect. 9) Images and real-time automatic flare detection from Kanzelhohe Observatory (Poetzi et al. 2015) located in Austria, pixel resolution is 1 arcsec and temporal resolution is 6 seconds (patrol mode).
- NST IR and optical data.
- Magnetograms and white light data from SDO
- Coronal magnetography maps obtained from imaging spectropolarimetry at 2.5-18 GHz from Expanded Owens Valley Array (EOVSA, solar-dedicated, ~13:30-02:00 UT); imaging spectropolarimetry at 1-8 GHz from VLA (coordinated observations are necessary), and other microwave imaging instruments all over the world.

- **Please also see the following science cases of the flare team H, which also proposes mosaics of Active Regions:**
- **SC-H1:** Interferometric observation of a flare @ 230 GHz
- **SC-H3:** Interferometric observation of a flare @ 100 GHz
- **SC-J4:** Thermal inhomogeneity of the solar corona

## L2.3 - Science Case SC-L3: Height-dependence of the brightness temperature spectrum in Active Regions

**Key Question SC-L3-KQ1: What is the local slope of the brightness temperature spectrum as function of observing frequency and thus sampled atmospheric height?**

**Key Question SC-L3-KQ2: What is the height scale of Active Region structure changes?**

Qualitatively similar to the science case SC-K2 for the Quiet Sun, the slope of the brightness temperature spectrum tells us about the height distribution of the plasma in chromosphere in Active Regions.

**Observational requirements**

- **Summary:**

  **Step 1.** Image a pre-selected AR @ four frequencies per band first in band 3 (~100 GHz) and then in band 6 (230 GHz) with ALMA.

  **Step 2.** Obtain spectra along a number of representative lines of sight (corresponding to various underlying photospheric components, selected based on the context observations) taken at various times.

  **Step 3.** Analyze local spectral slopes in both bands to estimate the local gradient of the plasma temperature and determine thermal structure of the solar chromosphere.

- **Target:** Active region close to solar disk center.
- **Array configuration:** Compact (C40-1 - C40-3).
- **Pointing/FOV:** Mosaic with 6x6 pointings and a FOV of 3arcmin x 3 arcmin (Band 3) or 1.3 arcm x 1.3 arcmin (Band 6).
- **Receiver:** The observation sequence should be carried out first with band 3 and then with band 6.
- **Temporal resolution:** 8-10min (the time it takes to complete the mosaic)
- **Duration:** 1-2 hours per band.

- **Please also see the following science cases of the flare team H, which also proposes mosaics of Active Regions:**
- **SC-H1:** Interferometric observation of a flare @ 230 GHz
- **SC-H3:** Interferometric observation of a flare @ 100 GHz
- **SC-J4:** Thermal inhomogeneity of the solar corona

### L2.4 - Science Case SC-L4: Temporal evolution of Active Regions

**SC-L4-KQ1: How does the brightness temperature spectrum varies with time for a given atmospheric component (umbra, penumbra, plage, facula, etc.)?**

Modern MHD models of the solar atmosphere suggest that the thermal distribution of the thermal plasma with height in the chromosphere is highly variable due to a plethora of dynamic processes, including shock waves. But it is yet unknown how exactly the thermal distribution with height varies in time and how the currently available MHD models capture at least the most essential features of the chromosphere.

**Observational requirements**

- **Summary:**
  **Step 1.** Image a pre-selected AR @ four frequencies per band first in band 3 (~100 GHz) and then in band 6 (230 GHz) with ALMA.
  **Step 2.** Obtain spectra along a number of representative AR brightness structures (corresponding to various underlying photospheric components, selected based on the context observations) taken with high cadence.
  **Step 3.** Study temporal variation of brightness spectrum for different brightness structures — how strong the instant height distributions of the atmosphere parameters can deviate from the standard stationary distribution—i.e., what is actual role of dynamic processes;

- **Target:** Active region close to solar disk center.
- **Array configuration:** Compact (C40-1-C40-3).
- **Pointing/FOV:** Single pointing centered on a selected feature (umbra, penumbra, plage).
- **Receiver:** The observation sequence should be carried out first in band 3 (~100 GHz) and then in band 6 (230 GHz) with ALMA. At least 4 spectral channels per band are needed.
- **Temporal resolution:** 2 s.
- **Duration:** 1 hour per band.

Please also see the following science cases of teams G, H and J, which also target Active Regions for high-cadence observations:
- **SC-G2:** Damping of Alfvén waves and associated heating
- **SC-G3:** Chromospheric flare oscillations
- **SC-G4:** Rapid oscillations in the chromosphere
- **SC-G8:** Quasi-periodic pulsations (QPPs)
- **SC-G9:** Sunspot oscillations
- **SC-H2:** Temporal and spatial evolution of a flare
- **SC-J1:** Chromospheric to coronal plasma response
- SC-J2: Dynamic nature of atmospheric heating
- SC-J3: Penumbral jets and bright dots
- SC-J4: Thermal inhomogeneity of the solar corona

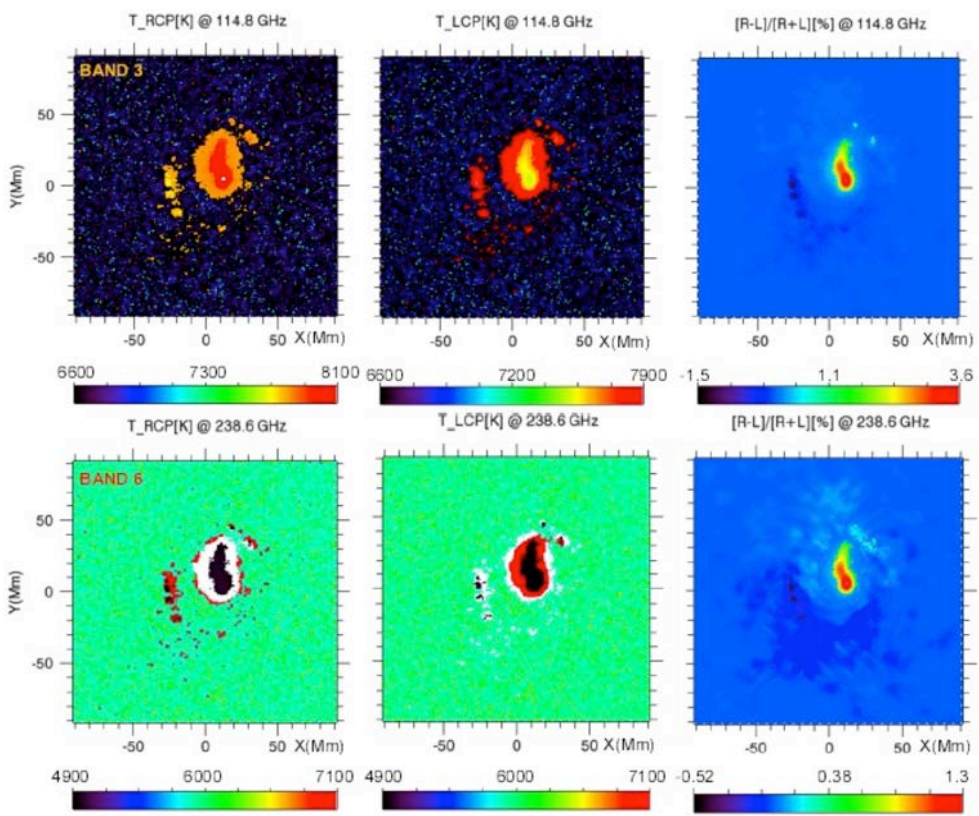

***Figure L.2:*** *Examples of simulated polarized emission maps (left circular polarization, right circular polarization and degree of polarization) in ALMA band 3 and band 6 for AR 12158 from Fig. L.1.*

## 12. SOLAR FLARES (TEAM H)

**List of team members contributing to this document:**
M. Battaglia, G. Gimenez de Castro, G. Fleishman, H. Hudson, P. Antolin, P. Simões, G. Doyle, A. Veronig, A. Benz, V. Nakariakov, J. Costa

**Lead author:** M. Battaglia *(marina.battaglia@fhnw.ch)*

### H1 - SUMMARY OF SCIENCE CASES AND KEY QUESTIONS

**Science Case SC-H1: Interferometric observation of a solar flare @ 230 GHz**

- **Key Question SC-H1-KQ1:** What is the (spatial) origin and morphology of solar flare emission at 230 GHz?

**Science Case SC-H2: Temporal and spatial evolution of a flare**

- **Key Question SC-H2-KQ1**: What is the temporal evolution of flare emission at 100 GHz?
- **Key Question SC-H2-KQ2**: Chromospheric manifestations of quasi-periodic pulsations?

**Science Case SC-H3: Interferometric observation of a solar flare @ 100 GHz**

- **Key Question SC-H3-KQ1:** What is the (spatial) origin and morphology of solar flare emission at 100 GHz?
- **Key Question SC-H3-KQ2:** What is the spectral shape of solar flare emission at 100 GHz?

**Beyond Cycle 4 - Science Case SC-H4: Measurement of a two-point flare spectrum**

*This science case is of high scientific interest but may only become feasible in a later observing cycle.*

- **Key Question SC-H4-KQ1:** What is the shape of the flare spectrum at 100-230 GHz?
- **Key Question SC-H4-KQ2:** What is the emission mechanism responsible for GHz flare emission?

### H2 - DETAILED DESCRIPTIONS OF SCIENCE CASES

#### H2.1 - Science Case SC-H1: Interferometric observation of a flare @ 230 GHz

**Key Question SC-H1-KQ1: What is the (spatial) origin and morphology of solar flare emission at 230 GHz?**

Solar flares are dynamic phenomena lasting from tens of seconds to minutes and evolving, both spectrally and spatially, in the process. They are observed over a wide wavelength range from X-rays to extreme ultraviolet (EUV), visible, infra-red (IR) and into the microwave (MW) and meter radio wavelength domain. However, not much is known about the origin and mechanism of solar flare emission at >100 GHz since this frequency range is still largely unexplored. High spatial resolution interferometry with ALMA, combined with context observations in microwaves (MW), H-alpha, the far-IR, UV, and X-rays is essential to determine the emission mechanism and height in the solar atmosphere. Despite their highly dynamic nature, new insight into flares can already be gained by imaging a flare at a specific time in its evolution and comparing this with context observation. Hence, **our first goal is to image any flare at 230 GHz (band 6).**

**Observing sequence:**

**Step 1:** Image a solar flare @ 230 GHz with ALMA.

**Step 2:** Compare location and timing of emission with emission at MWs, IR, white light, H-alpha, EUV and X-ray / gamma-ray wavelengths.

**Step 3:** Investigate morphology of flare, in comparison with context observations in other wavelengths and magnetograms to determine height of the emission in the solar atmosphere.

**Observational requirements**

- **Summary of observational strategy:** Band 6 interferometric observations, 6x6 mosaic of a pre-determined active region with full-Sun fast scanning with 3 TP dishes.
- **Target:** Pre-determined active region with highest flare potential, preferably near the limb.
- **Array configuration:** C40-2 (0.8 " resolution, 25" FOV @ 230 GHz → total FOV = <56")

- **Pointing:** Mosaic with 36 pointings.
- **Receiver:** Band 6
- **Temporal resolution:** 8-10min *(time for 1 complete mosaic within 5min of science operation plus 3-5min calibration)*
- **Duration:** Continuous observations for the full available time range where the Sun is above 40 degrees zenith angle (to prevent shadowing of antennas), but minimum 1 hour.

**Context observations**

- **H-alpha:** Images from the "H-alpha Telescope for Argentina" (HASTA) and images and real-time automatic flare detection from Kanzelhohe Observatory (Poetzi et al. 2015) located in Austria, pixel resolution is 1 arcsec and temporal resolution is 6 seconds (patrol mode).
- **Microwaves:** radio data at selected frequencies from 1 to 405 GHz
- **Full sun flux** from 1 to 14.5 GHz with 1 second temporal resolution from the Radio Solar Telescope Network (RSTN).
- **Full sun left / right polarization flux** at 45 and 90 GHz, and 10 ms temporal resolution from the POlarization Emission of Millimeter Activity at the Sun (POEMAS, Valio et al, 2013)
- **Flux at 212 and 405 GHz** with 5 ms temporal resolution from the Solar Submillimeter Telescope (SST, Kaufmann et al, 2008)
- **Microwaves:** imaging data
- **Imaging spectropolarimetry at 2.5-18 GHz** with 1s temporal resolution from Expanded Owens Valley Array (EOVSA, solar-dedicated, ~13:30-02:00 UT)
- **Imaging spectropolarimetry at 1-8 GHz** with 20 ms temporal resolution from VLA (coordinated observations are necessary)
- **Far-IR:** 30 THz images with 10 arcsec spatial resolution and 1 second temporal resolution (Kaufmann et al, 2013).
- **UV-EUV:** IRIS (imaging and spectroscopy, e.g. MgII=chromospheric, Si IV=lower transition region, FeXXI = upper transition region, corona), SDO/AIA (full Sun imaging)
- **X-rays:** Hinode, RHESSI (spectral imaging)
- **gamma-rays:** Fermi/GBM (spectroscopy)

> Please also see the following science cases, which propose to observe Active Regions and thus potential flares sites:
> - **SC-G2:** Damping of Alfvén waves and associated heating
> - **SC-G3:** Chromospheric flare oscillations
> - **SC-G4:** Rapid oscillations in the chromosphere
> - **SC-G8:** Quasi-periodic pulsations (QPPs)
> - **SC-G9:** Sunspot oscillations
> - **SC-L1:** The brightness temperature distribution in active regions and sunspots
> - **SC-L2:** The dependence of Active Region brightness temperatures on magnetic field **strength**
> - **SC-L3:** Height-dependence of the brightness temperature spectrum in Active Regions
> - **SC-L4:** Temporal evolution of Active Regions
> - **SC-J1:** Chromospheric to coronal plasma response
> - **SC-J2:** Dynamic nature of atmospheric heating
> - **SC-J3:** Penumbral jets and bright dots
> - **SC-J4:** Thermal inhomogeneity of the solar corona

---

**H2.2 - Science Case SC-H2: Temporal and spatial evolution of a flare**

**Key Questions: SC-H2-KQ1**: **What is the temporal evolution of flare emission at 100 GHz? SC-H2-KQ2: Chromospheric manifestations of quasi-periodic pulsations?**

The flare duration can last from the order of minutes for small flares up to hours for very large flares. Over this time, the location and intensity of emission at different wavelengths is known to change considerably. The first objective of this second science case is to observe the time-evolution of the flaring emission at 100 GHz (band 3) and compare it with the evolution observed at other wavelengths. As a second objective, we will investigate the presence and nature of flare-generated chromospheric oscillations. Oscillations (with periods from a fraction of a second to tens of minutes) over a vast wavelength range are frequently observed in solar flares, most commonly at hard X-rays and microwaves, and more recently at soft X-rays. Oscillations with a periodicity of the order of tens of seconds (originating from the chromosphere) have been observed in the UV in both, solar flares and stellar flares, suggesting that such oscillations may be an integral part of the chromospheric flare process. With ALMA's ability to observe the chromospheric region, and its sub-arcsec spatial resolution, we wish to observe 1 or 2 solar flares with a time resolution of 1-2 sec, simultaneously with context instruments that observe the lower atmosphere, e.g. IRIS in Mg II (chromosphere) and Si IV (lower transition region), and the corona, e.g. Hinode and EOVSA. The larger FOV of band 3 allows for a sit-and-stare observational strategy, increasing the temporal resolution.

**Observing sequence:**

**Step 1:** Observe a flare at high time resolution @ 100 GHz.

**Step 2:** Compare evolution of source-integrated intensity with context information at other wavelengths.

**Step 3:** Analyse spatial evolution and compare with context observations.

**Step 4:** Determine the horizontal and vertical spatial fine structure and variation of quasi-periodic variations of the 100 GHz emission, and compare it with chromospheric, transition region and coronal context observations.

**Observational requirements**

- **Summary of observational strategy:** Band 3 interferometric observations, sit-and-stare (single pointing) of a pre-determined active region with full-Sun fast scanning with 3 TP dishes.
- **Target:** Pre-determined active region with highest flare potential, preferably near the limb.
- **Array configuration:** C40-2 (1.8 " resolution, 58" FOV @ 100 GHz)
- **Pointing:** Single.
- **Receiver:** Band 3
- **Temporal resolution:** 1-2 seconds
- **Duration:** Continuous observations for the full available time range where the Sun is above 40 degrees zenith angle (to prevent shadowing of antennas), but minimum 1 hour.

**Context observations**

Same as SC-H1.

---

**Please also see the following science cases, which propose to observe Active Regions and thus potential flares sites:**

- **SC-G2:** Damping of Alfvén waves and associated heating
- **SC-G3:** Chromospheric flare oscillations
- **SC-G4:** Rapid oscillations in the chromosphere
- **SC-G8:** Quasi-periodic pulsations (QPPs)
- **SC-G9:** Sunspot oscillations
- **SC-L1:** The brightness temperature distribution in active regions and sunspots
- **SC-L2:** The dependence of Active Region brightness temperatures on magnetic field **strength**
- **SC-L3:** Height-dependence of the brightness temperature spectrum in Active Regions
- **SC-L4:** Temporal evolution of Active Regions
- **SC-J1:** Chromospheric to coronal plasma response
- **SC-J2:** Dynamic nature of atmospheric heating
- **SC-J3:** Penumbral jets and bright dots
- **SC-J4:** Thermal inhomogeneity of the solar corona

---

## H2.3 - Science Case SC-H3: Interferometric observation of a flare @ 100 GHz

**Key Questions: SC-H3-KQ1:** What is the (spatial) origin and morphology of solar flare emission at 100 GHz? **SC-H3-KQ2:** What is the spectral shape of solar flare emission at 100 GHz?

Microwave emission from flares is mainly produced by accelerated electrons through the gyrosynchrotron mechanism, with the spectrum decreasing in intensity at higher frequencies. Towards the higher frequencies thermal free-free emission becomes more important, especially during the decay phase of flares. Above 100 GHz, a new spectrum component has been identified, with increasing intensity with frequency. However, its origin remains a mystery. Any of these three spectral components can play a dominant role in the spectrum above 100 GHz. The 4-subchannel spectrum in band 3 will provide valuable information about the nature of the emission mechanism simply by showing the slope of the spectrum: positive (increasing), negative (decreasing) or flat. Having this information at different locations of the flare with interferometric observations, and comparing with data from the spatially resolved context observations would provide further clues about the nature of the emitting mechanism. Furthermore, obtaining the spectrum at different phases of a flare is also of extreme importance, capturing the impulsive and gradual phases, when the effects from non-thermal electrons or heated plasma are respectively more relevant.

**Observing sequence:**

**Step 1:** Image a solar flare @ 100 GHz with ALMA.

**Step 2:** Identify the slope of the 4-subchannel spectrum (positive, negative or flat) at different locations/times.

**Step 3:** Compare location and timing of emission with emission at MWs, IR, white light, H-alpha, EUV and X-ray / gamma-ray wavelengths to infer the thermal or non-thermal nature of the 100 GHz sources

**Observational requirements**

- **Summary of observational strategy:** Band 3 interferometric observations, 6x6 mosaic of a pre-determined active region with full-Sun fast scanning with 3 TP dishes.

- **Target:** Pre-determined active region with highest flare potential, preferably near the limb.
- **Array configuration:** C40-2 (1.8 " resolution, 58" FOV @ 100 GHz → total FOV = 90 " for 6x6 mosaic)
- **Pointing/FOV:** 6x6 mosaic.
- **Receiver:** Band 3
- **Temporal resolution:** 8-10min *(time for 1 complete mosaic within 5min of science operation plus 3-5min calibration)*
- **Duration:** Continuous observations for the full available time range where the Sun is above 40 degrees zenith angle (to prevent shadowing of antennas), but minimum 1 hour.

**Context observations:**
Same as SC-H1.

**Please also see the following science cases, which propose to observe Active Regions and thus potential flares sites:**

- **SC-G2:** Damping of Alfvén waves and associated heating
- **SC-G3:** Chromospheric flare oscillations
- **SC-G4:** Rapid oscillations in the chromosphere
- **SC-G8:** Quasi-periodic pulsations (QPPs)
- **SC-G9:** Sunspot oscillations
- **SC-L1:** The brightness temperature distribution in active regions and sunspots
- **SC-L2:** The dependence of Active Region brightness temperatures on magnetic field **strength**
- **SC-L3:** Height-dependence of the brightness temperature spectrum in Active Regions
- **SC-L4:** Temporal evolution of Active Regions
- **SC-J1:** Chromospheric to coronal plasma response
- **SC-J2:** Dynamic nature of atmospheric heating
- **SC-J3:** Penumbral jets and bright dots
- **SC-J4:** Thermal inhomogeneity of the solar corona

**H2.4 - Beyond cycle 4 -   Science Case SC-H4: Measurement of a two-point flare spectrum**

**Technical requirements / necessary capabilities for optimal science output:**

- Simultaneous observations at two frequencies (band 3 & 6)
- Temporal resolution of < 10s
- Possibility to re-point in real-time, i.e. fast scanning of a (or several) active regions with change to single pointing at part of specific active region when a flare happens.

## 13. PROMINENCES (TEAM I)

**List of contributing team members:**
P. Antolin, J.L. Ballester, R. Brajsa, S. Gunár, N. Labrosse, B. Schmieder, M. Temmer

**Lead author:** *N. Labrosse (Nicolas.Labrosse@glasgow.ac.uk)*

### I1 - SUMMARY OF SCIENCE CASES AND KEY QUESTIONS

Solar prominences (also known as filaments when they are viewed against the solar disc) are large structures located in the solar corona and one of the most dramatic illustrations of how the magnetic field shapes the atmosphere of the Sun. They are regions of cool, dense plasma suspended in the prominence magnetic field within the hot and rarefied coronal environment. While the large-scale prominence structures remain relatively stable for long periods of time, their finest features with dimensions on scales of a few hundred kilometres undergo dynamical changes on timescales of minutes. The prominence plasma is also thermodynamically very active. Their substructures change shape continuously, and both small or large portions of a prominence can suddenly heat up in response to heating mechanisms (a phenomenon called *prominence activation*). This can possibly lead to the prominence plasma disappearing altogether from the chromospheric temperature range.

Occasionally, the strongly sheared prominence magnetic field becomes unstable. Prominences can then suddenly loose equilibrium and erupt, sending huge amounts of magnetised plasma in the interplanetary space. These eruptions are closely related to two of the most violent phenomena found in the solar system, namely, Solar Flares and Coronal Mass Ejections (CMEs).

At high resolution, prominences are composed of fine sub-arcsecond threads which may reflect elementary magnetic field structures. The radiation emitted by the cool prominence plasma in wavelengths in which high resolution can be achieved may allow the prominence magnetic field to be visualised, and be used to study important physical processes such as coronal heating.

Understanding the formation of solar prominences, their spatial and temporal evolutions, and subsequently the origin of Solar Flares and CMEs, is therefore of the utmost importance. Observations with the highest possible spatial and temporal resolution will enable us to track the changes of the prominence plasma and magnetic field, and to describe in detail the process of prominence evolution that may lead to an eruption.

The temperatures of the cool dense prominence plasma are in the region of $10^4$ K (chromospheric temperatures). In the optical and UV wavelength ranges, most of the spectral lines used for observations and for the diagnostics of the plasma are optically thick and thus have to be treated by radiative transfer. On the other hand, at sub-millimetric wavelengths the prominence plasma will be optically thin, and observations much easier to interpret. The primary contribution of ALMA to the understanding of solar prominences is its unique capability to measure the temperature of the cool prominence plasma with high spatial resolution, and follow its evolution over time.

We identify the following two strong Science Cases underpinned by several key questions that we want to address with ALMA:

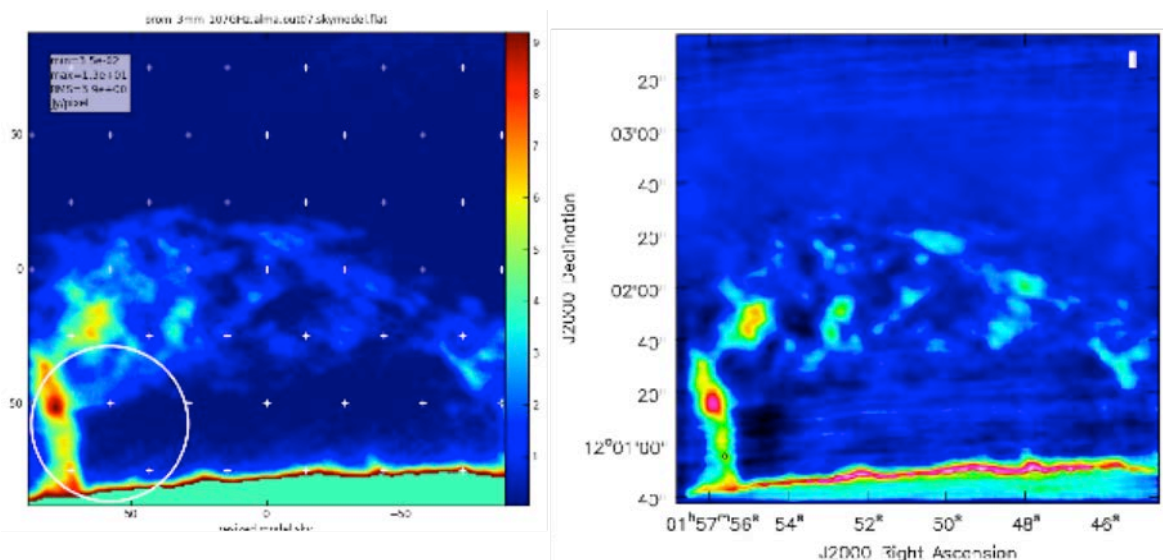

*Figure I.1: Simulated ALMA mosaic at 100 GHz. Left: Ideal brightness temperature model. Right: Simulated ALMA mosaic. Taken from Heinzel et al. (2015).*

**Science Case SC-I1: The thermal structure of solar prominences at millimetre wavelengths**

- **Key Question SC-I1-KQ1:** What is the fine-scale thermal structure of solar prominences and filaments at high spatial resolution in their main body and in the prominence-corona transition region?
(Cycle 4 priority: high; scientific priority: high; Technical feasibility for Cycle 4: feasible)
- **Key Question SC-I1-KQ2:** How does the prominence plasma react to various heating processes?
(Cycle 4 priority: high; scientific priority: high; Technical feasibility for Cycle 4: feasible)
- **Key Question SC-I1-KQ3:** How is the dynamics of the plasma related on small scales and on large scales to the structure in temperature of solar prominences?
(Cycle 4 priority: high; scientific priority: high; Technical feasibility for Cycle 4: challenging)

**Science Case SC-I2: The spatial structure of solar prominences at millimetre wavelengths**

- **Key Question SC-I2-KQ1:** How is the fine-scale structure of prominences shaped by the magnetic field? (Cycle 4 priority: high; scientific priority: high; Technical feasibility for Cycle 4: feasible)
- **Key Question SC-I2-KQ2:** How do Active Region and Quiet Sun prominences differ in millimetre wavelengths?
(Cycle 4 priority: low; scientific priority: low; Technical feasibility for Cycle 4: feasible)

We refer to Labrosse et al. (2010, Space Sci. Rev. 151, 243) and the recent book by Vial & Engvold (2015, Astrophysics and Space Science Library, Vol. 415) for an extended discussion of open questions on solar prominences. More recently, the study by Heinzel et al. (2015, Solar Phys. 290, 1981) provides an insight on the visibility of prominence fine structures at radio millimetre wavelengths with ALMA.

## I2 - DETAILED DESCRIPTIONS OF SCIENCE CASES

### I2.1 - Science Case SC-I1: The thermal structure of solar prominences at millimetre wavelengths

The questions we propose to address in relation to SC-I1-KQ1 are:
- What is the fine-scale thermal structure of solar prominences and filaments at high spatial resolution in their main body and in the prominence-corona transition region?
- How cold can temperatures get in prominences?
- What is the spatial distribution of such temperatures within the prominence under different viewing angles?
- What is the timescale and range of the change in temperature?

The questions we propose to address in relation to SC-I1-KQ2 are:
- **How does the prominence plasma react to various heating processes?**
- **What are the most significant thermal / non-thermal processes in prominences?**

The questions we propose to address in relation to SC-I1-KQ3 are:
- **What is the nature of the different types of oscillations in solar prominences?**
- **How do filament and prominence eruptions look like at ALMA wavelengths?**

ALMA will give us very precise temperature estimates since most of the contribution to the flux is from thermal processes. The high temporal cadence of ALMA, combined with other instrumentation such as SDO, IRIS, and Hinode (from space), and ground-based observatories (e.g. high-resolution H-alpha observations, NST, assisted by Kanzelhöhe Observatory and its automatic algorithm applied on H-alpha data for the detection of filaments to obtain their position and shape parameters), should be able to provide unique insights into the thermal evolution of prominence plasma subject to heating mechanisms such as waves or field restructuring from slow footpoint stressing. Changes in the thermodynamic parameters of filaments/prominences should be studied with respect to the underlying magnetic field (e.g., vortices, tornado structure). Note that SDO/HMI provides high resolution magnetic field data and will be available all the time.

Important observables that can be obtained with ALMA are:

- height dependence of thermal emission (off-limb study)

- wave power versus period
- flow magnitudes
- timescale / range of temperature changes
- onset of prominence activation

Concentrating on the thermal processes (non-thermal processes will be studied in Cycle 5), we will investigate prominence heating mechanisms. At high cadence we will aim at detecting turbulence in the power spectrum. For identifying prominence activation and heating mechanisms, help from coordinated observations with other instruments will be useful. It is very important to understand local or non local heating and cooling processes in prominences since they can affect wave damping or amplification. The reported oscillatory periods in small amplitude oscillations have been classified in four groups (Lin, 2011, Space Science Reviews, 158, 237): very short periods (≤ 1 min), short (1 min – 20 min), intermediate (20 - 40 min) and long periods (40 - 100 min). Depending on the types of oscillations that may be observed (period, amplitude), the relevance of each potential damping mechanism could be assessed by comparing the spatial and time scales produced by each of them (as inferred from the interpretation of oscillations in terms of MHD waves) with those obtained from observations (Arregui & Ballester, 2011, Space Science Reviews, 158, 169).

Answering these questions will provide very powerful constraints for the modelling of the energy balance in prominences and for our understanding of how they form, evolve, and erupt.

**Observational requirements**

- **Observing mode:** Interferometry + Total Power mode (full-disk fast scanning with 3 TP antennas, 2400" x 2400" FOV)
- **Target:** Prominence off-limb if possible and/or filament on disk
- **Array configuration:** 12m + 7m + TP in compact mode (C40-2)
    - **Spatial resolution:** < 1 arcsec preferable
    - **Range of scales:** from resolution limit up to field of view
- **Pointings/FOV:** mosaic with 6x6 pointings
- **Receiver:** First band 3, then band 6
    - **Sensitivity:** better than 100 K
    - **Flux calibration accuracy:** <5%
- **Temporal resolution:** ~8 min *(the time it takes to complete a mosaic plus calibration)*
- **Duration:** 2 x 95 min per band (~6 hours in total incl. 4 times initial calibration)

**Proposed observation**

Initially, we assumed calibration times of 5min and max science observation duration of 5min. These numbers have now been changed to 10min and 3min, respectively, which increases the number of science observation steps in a 1 hour window and maximises scientific data return.

We suggest two runs of the following sequence:

1. Receiver setup for **band 3** (a few min)
2. Bandpass, sideband separation, flux, pointing calibration (25-30 min)
3. Band 3: 5 sequences of a) science observation (10 min) and b) intermediate calibration (3 min) → 65 min
4. Receiver setup for **band 6** (a few min)
5. Bandpass, sideband separation, flux, pointing calibration (25-30 min)
6. Band 6: 5 sequences of a) science observation (10 min) and b) intermediate calibration (3 min) → 65 min

A single run through steps 1 to 6 would be desirable for the test observations in December 2015.

---

**I2.2 - Science Case SC-I2: The spatial structure of solar prominences at millimetre wavelengths**

The questions we propose to address in relation to SC-I2-KQ1 are:
- What is the fine-scale structure of solar prominences at millimetre wavelengths?
- Can prominence threads be distinguished?
- What do the observed fine structures tell us about the magnetic field?

The question we propose to address in relation to SC-I2-KQ2 is: **How do Active Region and Quiet Sun prominences differ in millimetre wavelengths?**

ALMA will give us a very high spatial resolution view of the fine structure of limb prominences and filaments. The high temporal cadence of ALMA, combined with other instrumentation such as SDO, IRIS, and Hinode (from space), and ground-based observatories (e.g. high-resolution H-alpha observations, NST, assisted by Kanzelhöhe Observatory and its automatic algorithm applied on H-alpha data for the detection of filaments to obtain their position and shape parameters), should be able to provide unique insights into the temporal behaviour of small-scale features. Changes in the fine structure of filaments/prominences should be studied with respect to the underlying magnetic field (e.g., vortices, tornado structure). Note that SDO/HMI provides high resolution magnetic field data and will be available all the time. Through coordination with other instruments probing hotter temperatures, more precise estimates of the prominence-to-corona transition region sizes may be achieved.

Important observables that can be obtained with ALMA are:

- the distribution and sizes of the fine structures at ALMA wavelengths,
- the dynamics and temporal evolution of the fine structures,
- the filling factor, corresponding to the fraction occupied by the fine structures.

We will investigate the mechanisms responsible for the observed motions of prominence fine structures. The distributions and sizes of fine structures will help us to constrain prominence models and better understand the interplay between fine structure and magnetic field. High-cadence observations will allow us to study the evolution of their characteristics in different regions of the prominence, with a particular focus on so-called tornadoes often detected in the legs / barbs of prominences.

Answering these questions will allow us to advance models of the fine structure of prominences and filaments and of their temporal evolution, for instance by constraining transport coefficients in the partially ionised plasma of prominences.

**Observational requirements**

- **Observing mode:** Interferometry + Total Power mode (full-disk fast scanning with 3 TP antennas, 2400" x 2400" FOV)
- **Targets: (1) Quiet Sun** prominence off-limb or filament on disk. (2) Active Region prominence off-limb or filament on disk.
- **Array configuration:** 12m + 7m + TP in compact mode (C40-2)
    - **Spatial resolution:** < 1 arcsec preferable
    - **Range of scales:** from resolution limit up to field of view
- **Pointings/FOV:** first 6 mosaics with 6x6 pointings, then a 150-point mosaic
- **Receiver:** First band 3, then band 6
    - **Sensitivity:** better than 100 K
    - **Flux calibration accuracy:** <5%
- **Temporal resolution:** 10 min for the small mosaic, 26 min for the large mosaic *(i.e. the time it takes to complete a mosaic plus calibration)*
- **Duration:** 30 min initial calibration + 39 min for six 6x6 mosaics + 26 min for one 150-p. mosaic, resulting in 95 min per band and per target (3:10 per target).

**Proposed observation**

We propose the following sequence for each target (i.e., a Quiet Sun prominence and an Active Region prominence).

1. Receiver setup for **band 3** (a few min)
2. Bandpass, sideband separation, flux, pointing calibration (25-30 min)
3. Band 3: six 6x6 mosaics (3 x (10+3) min) → 39 min
4. Band 3: 150-point mosaic of the same target (26 min)
5. Receiver setup for **band 6** (a few min)
6. Bandpass, sideband separation, flux, pointing calibration (25-30 min)
7. Band 6: six 6x6 mosaics (3 x (10+3) min) → 39 min
8. Band 6: 150-point mosaic of the same target (26 min)

A single run through steps 1 to 6 would be desirable for the test observations in December 2015.

## 14. OSCILLATIONS AND WAVES (TEAM G)

**List of team members contributing to this document:**
P. Antolin, J. Ballester, G. Doyle, B. Fleck, M. Loukitcheva, V. Nakariakov, R. Oliver, R. Soler, T. Van Doorsselaere, S. Wedemeyer, T. Zaqarashvili

**Lead author:** B. Fleck *(bfleck@esa.nascom.nasa.gov)*

### G1 - SUMMARY OF SCIENCE CASES AND KEY QUESTIONS

The solar atmosphere is a stratified medium permeated by magnetic field with a complicated topology, which makes it very dynamic in nature. All types of region on the Sun exhibit a large variety of oscillations and propagating waves. Well known are the 5-min and 3-min oscillations, which are most dominant in the photosphere and chromosphere in quiet Sun regions, respectively. Propagating waves, which predominantly are excited in the lower layers and propagate upwards, occur in different wave modes such as purely acoustic or Alfvén but also undergo mode conversion, in particular when encountering magnetic field structures. The observed properties of oscillations and waves allow to determine the properties of the affected atmospheric layers, e.g. the stratification of gas pressure and temperature and the magnetic field. Furthermore, waves carry energy from the lower layers into the upper layers, where it is dissipated and thus contributes to the heating of the solar chromosphere and corona. Studies of oscillations and waves with ALMA, in particular given ALMA's ability to serve a linear thermometer sampling many different heights, provide a wealth of diagnostic opportunities, which will lead to a better understanding of the structure and dynamics of the solar atmosphere.

**Science Case SC-G1: 2.1 Acoustic waves**

- **Key question SC-G1-KQ1:** What are the propagation characteristics of acoustic waves in the chromosphere?
- **Key question SC-G1-KQ2:** How much energy do they transport into the higher atmosphere?

**Science Case SC-G2: Damping of Alfvén waves and associated heating**

- **Key question SC-G2-KQ1:** What is the role of overdamped high-frequency Alfvén waves in chromospheric heating?
- **Key question SC-G2-KQ2:** Can we indirectly estimate physical conditions of the plasma (e.g., transport coefficients) by observing damped waves?

**Science Case SC-G3: Chromospheric flare oscillations**

- **Key question SC-G3-KQ1:** What is the spatial location and extent of chromospheric oscillations in solar flares?
- **Key question SC-G3-KQ2:** What is the height variation of these oscillations, compared with say Mg II or Si IV?

**Science Case SC-G4: Rapid oscillations in the chromosphere**

- **Key question SC-G4-KQ1:** Where and how are rapid oscillations excited in the chromosphere?
- **Key question SC-G4-KQ2:** What chromospheric structures could be the driver for the rapid oscillations in the solar corona?

**Science Case SC-G5: High frequency waves in the chromosphere**

- **Key question SC-G5-KQ1:** Where are high-frequency MHD waves excited?
- **Key question SC-G5-KQ2:** What is the energy flux of these waves in the chromosphere?

**Science Case SC-G6: Seismology of chromospheric fibrils**

- **Key question SC-G6-KQ1:** Do chromospheric fibrils support quasi-periodic propagating waves?
- **Key question SC-G6-KQ2:** If detected, can the spatial and/or time signatures of these waves be used for the seismology of chromospheric fibrils?

ALMA OBSERVATIONS OF THE SUN IN CYCLE 4 AND BEYOND**Science Case SC-G7: Short period oscillations in filament threads**
- **Key question SC-G7-KQ1:** Detection of very short period oscillations in filament threads

**Science Case SC-G8: Quasi-periodic pulsations (QPPs)**
- **Key question SC-G8-KQ1:** Fine spatial and time structure of phases, periods and power of QPP in flares

**Science Case SC-G9: Sunspot oscillations**
- **Key question SC-G9-KQ1:** Fine spatial structure of umbral oscillations

## G2 - DETAILED DESCRIPTIONS OF SCIENCE CASES

### G2.1 - Science Case SC-G1: Acoustic waves

**Key questions: What are the propagation characteristics of acoustic waves in the chromosphere? How much energy do they transport into the higher atmosphere?**

Waves and oscillations are interesting not only from the point of view that they can propagate energy into the chromosphere and dissipate that energy to produce non-radiative heating, but they also carry information about the structure of the atmosphere in which they propagate. Since the sound speed is a slowly varying function of temperature in the chromosphere, pressure disturbances with frequencies above the acoustic cut-off propagating upwards in the chromosphere provide information about the difference of the formation height of the accompanying temperature and velocity perturbations, or, reversely, if the difference of formation heights is known, allow inferences about the phase speed of the waves and hence about the wave mode.

Several authors (Mein 1971; Mein & Mein 1976; Lites et al. 1982; Fleck & Deubner 1989) who studied the wave propagation behavior in the chromosphere by analyzing time series of the Ca II infrared triplet lines at 854.2 nm and 849.8 nm found zero phase differences between the Doppler shifts of the two lines. The vanishing phase lag between the two lines was interpreted as evidence of a standing wave pattern in the chromosphere caused by total reflection of the waves at the transition region. This conclusion was based on the assumption that the Doppler signal of the two Ca II infrared lines is formed at different heights in the chromosphere, an assumption that was later challenged (Skartlien et al. 1994). Fleck et al. (1994a, b) extended the earlier work of Fleck & Deubner (1989) and included the Ca II K and He I 1083 nm lines together with the Ca II 854.2 nm line in their observations. It is generally believed that that He I 10830 nm line is formed high up in the chromosphere just below the transition region and that the formation height of Ca II K3 also is different from that of the Ca II infrared triplet (Skartlien et al. 1994). The phase difference between these three lines, however, still showed nearly uniform value close to 0°, especially at frequencies above the cutoff frequency, providing further evidence of a dominating non-propagating component of the chromospheric wave field.

According to the dispersion relation of acoustic-gravity waves in a stratified atmosphere, propagating sound waves display a linear increase of the phase difference with frequency. Interestingly, the analysis of recent IRIS observations of the Mg II h and k line also shows vanishing differences between the $k_3$, $k_{2v}$ and $k_{2r}$ features, which are believed to form at different heights. Why is the apparent phase speed of high frequency acoustic waves so high? Are these results misleading because of complex radiation transfer effects in these optically thick lines?

Loukitcheva et al. (2004, 2006, 2008), and White et al. (2006) have demonstrated the feasibility of measuring chromospheric oscillations in the mm range. Multi-wavelength time series of ALMA observations of the temperature fluctuations of inter-network oscillations should allow travel time measurements between different heights as these disturbances propagate through the chromosphere and thus should finally settle the long-standing question about the propagation characteristics of acoustic waves in the chromosphere.

**Observational requirements**
- **Target:** Quiet Sun regions near disk center
- **Antenna configuration:** one run in C40-1, one run in C40-2
- **Pointing/FOV:** single *(max. FOV possible for given antenna configuration)*
- **Receiver:** in both antenna configurations one run each in band 3 (100 GHz), one run in band 6 (230 GHz), and - if possible - multi-band runs using bands 3 and 6 simultaneously. The reason for the latter is that the expected height coverage in single band runs is rather small (< 100 km; see Fig. 7 of Loukitcheva et al. 2004). As a result, the expected phase differences will be very small and difficult to measure, in particular if the time series are limited in length.

http://ssalmon.uio.no                                                                                                                          41

Cross-band time series would allow probing significantly different atmospheric layers (300-500 km height difference), allowing much easier and more precise measurements of travel time differences. This will require detailed simulations of the sub-array division to achieve similar spatial resolution and uv-coverage in both bands.

- **Temporal resolution:** 2 s
- **Duration:** 30 min *(plus calibration interruptions; those as short as possible)*

**Please also see the science cases by Team K, which focus on the Quiet Sun.**

## G2.2 - Science Case SC-G2: Damping of Alfvén waves and associated heating

**Key questions: What is the role of overdamped high-frequency Alfvén waves in chromospheric heating? Can we indirectly estimate physical conditions of the plasma (e.g., transport coefficients) by observing damped waves?**

The physical processes responsible for the transport of energy from the solar interior and its dissipation in the atmospheric plasma are under intense research (see, e.g., Parnell & De Moortel 2012). One of the mechanisms that has been proposed to explain the transport and dissipation of energy involves the propagation and damping of Alfvénic waves (see the recent review by Arregui 2015 and references therein). Observations indicate that Alfvén waves are ubiquitous in the solar atmosphere and can have the required energy to heat the upper layers (see, e.g., Cargill & de Moortel 2011; McIntosh et al. 2011; Hahn & Savin 2014; Jess et al. 2015). While the overwhelming presence of the waves is demonstrated by the observations, the physics behind the damping of the waves and the deposition of wave energy into the plasma remains poorly explored. Numerical simulations have shown that partial ionization effects have a strong impact on chromospheric dynamics (see, e.g., Martínez-Sykora et al. 2012; Leake et al. 2014). Dissipation produced by ion-neutral collisions may play a crucial role in the release of magnetic energy in the form of heat (Khomenko & Collados 2012). The consideration of partial ionization effects must necessarily be done beyond the classic MHD description of plasma dynamics. The use of correct and accurate transport coefficients that govern basic collisional phenomena in the plasma is, therefore, essential for the realistic modeling of the chromosphere.

Using simplified chromospheric models, theoretical estimations of the damping rate of Alfvén waves due to ion-neutral collisions indicate that waves with high frequencies can be critically damped, i.e., overdamped (see, e.g., De Pontieu et al. 2001; Khodachenko et al. 2004; Leake et al. 2005; Soler et al. 2015). The requirement for overdamping is that the wave frequency approaches the ion-neutral collision frequency of the plasma, which is a function of the local physical conditions and depends on height in the chromosphere. The energy carried by these overdamped waves can be efficiently deposited in the plasma as a result of the strong dissipation. Computations of the heating obtained from numerical simulations including partial ionization effects (see Goodman 2011; Song & Vasyliūnas 2011; Russell & Fletcher 2013; Tu & Song 2013) confirm that dissipation of Alfvén wave energy can provide a sustained heating over time that could be sufficient to compensate the chromospheric radiative losses at low altitudes.

ALMA, with its 2 s cadence, is an excellent instrument to observe those high-frequency waves in the chromosphere and so to explore their relevance for chromospheric heating. On the one hand, the detection of high-frequency Alfvén waves would offer a tool to test the applicability of MHD and multi-fluid theories. On the other hand, it could be possible to indirectly infer physical conditions and/or transport coefficients that govern basic collisional phenomena (e.g., ion-neutral collisions) in the plasma. Using different spectral windows, the efficiency of the damping as a function of the wave frequency could be tracked at different heights. Then, observations could be compared with the Alfvén wave damping predicted by theoretical and numerical modeling.

**Observational requirements**

High cadence observations are needed for the detection of high-frequency waves. High spatial resolution is also needed, as the associated wavelengths are very small.

- **Target:** zones of intense magnetic field, chromospheric network, active region plage.
- **Antenna configuration:** C40-3 or C40-2.
- **Pointing/FOV:** single (FOV: maximum possible)
- **Receiver:** Band 6 (230 GHz) (if possible, also one run in Band 3 (100 GHz)).
- **Temporal resolution:** 2 s
- **Duration:** 10 min

**Please also see the science cases by Team K, which focus on the Quiet Sun.**

### G2.3 - Science Case SC-G3: Chromospheric flare oscillations

**Key Questions: What is the spatial location and extent of chromospheric oscillations in solar flares? What is the height variation of these oscillations, compared with say Mg II or Si IV?**

Oscillations over a vast wavelength range are frequently observed in solar flares. For example, oscillations in the range of 8.5 sec (originating from the chromosphere) were observed in the UV in a solar flare (van Doorsselaere et al., 2011). Similarly, in stellar flares oscillations in the range of 20-30 sec were observed in the UV, suggesting that such oscillations maybe an integral part of the chromospheric flare process. In both cases, the sausage mode was the preferred interpretation (Welsch et al. 2006). With ALMA's ability to observe the chromospheric region, we wish to observe 1 or 2 solar flares with a time resolution of 2 sec, simultaneous with space-based instruments, e.g. IRIS in Mg II (chromosphere) and Si IV (lower transition region) and Hinode (corona).

**Observational requirements**

- **Target:** the trailing spot in an active region
- **Antenna configuration:** C40-1
- **Pointing/FOV:** single *(FOV: maximum possible)*
- **Receiver:** Band 6 (230 GHz)
- **Temporal resolution:** s s
- **Duration:** 30 min *(plus calibration interruptions; those as short as possible)*

### G2.4 - Science Case SC-G4: Rapid oscillations in the chromosphere

**Key questions: Where and how are rapid oscillations excited in the chromosphere? What chromospheric structures could be the driver for the rapid oscillations in the solar corona?**

Williams et al. (2001) observed 6s periodicities in a coronal active region during a total solar eclipse. In this Science Case, we aim at detecting similar periodicities in the chromosphere, in order to investigate where and how they are excited. Outside of a solar eclipse, such short period waves are still not accessible to space-bourne instruments such as SDO/AIA or IRIS. Only with radio instruments, such short periods can be detected. However, with current instruments such as the Nobeyama Radio-Heliograph, the spatial resolution is too coarse to detect these fast oscillations. Thus, ALMA is ideally suited to detect these high-frequency oscillations, offering both an excellent time resolution and spatial resolution.

The rapid oscillations may be related to fast sausage modes, for which short periods are expected (e.g. Inglis et al. 2009). The detection of these waves would offer thus an extra seismological tool for the inference of plasma parameters in the solar chromosphere.

**ALMA OBSERVATIONS OF THE SUN IN CYCLE 4 AND BEYOND**## Observational requirements

For the detection of the rapid oscillations, we need high cadence, high resolution observations in active region plages. ALMA should focus on one field-of-view in a sit-and-stare mode for the full observation window (10 min). The highest possible cadence is necessary to capture the short period oscillations, i.e. 2 s in Cycle 4. Given the small periods, we also expect small wavelengths, and thus observations in C40-3 with at 230GHz are preferable. C40-2 could be used if the C40-3 is not available.

- **Target:** active region plage, sit-and-stare interferometric mode
- **Antenna configuration:** C40-3 in band 6 (230 GHz), or, if C40-3 is not available, C40-2.
- **Pointing/FOV:** single *(FOV: maximum possible)*
- **Receiver:** Band 6 (230 GHz) (and, if possible, also one run in Band 3 (100GHz)).
- **Temporal resolution:** 2 s
- **Duration:** 10 min

### G2.5 - Science Case SC-G5: High frequency waves in the chromosphere

**Key questions: Where are high-frequency MHD waves excited? What is the energy flux of these waves in the chromosphere?**

Solar granulation excited a broad spectrum of magnetohydrodynamic waves in magnetic flux tubes in the solar photosphere. These waves transport energy upwards and heat chromosphere/corona by the dissipation through different mechanisms. Low-frequency waves with periods of > 2-3 min are well studied by different space missions and ground based telescopes, but the observed energy flux of the waves apparently is not enough to compensate for the chromospheric and coronal energy losses.

On the other hand, high-frequency waves with periods of < 1 min are not well studied, due to insufficient temporal resolution, though they sometimes appear on the disk (Wunnenberg et al. 2002, de Wijn et al. 2005, Jess et al. 2007, with periods of > 20 s), at the limb (Zaqarashvili et al. 2007, Morton et al. 2014, with periods > 20 s) and in solar eclipse observations (Williams et al. 2002, Katsiyannis et al. 2003, with periods of 4-7s). These waves may provide (at least part of) the energy required for heating the quiet chromosphere and corona. ALMA, with its 1 s cadence, is an excellent instrument to observe waves with periods of 5-60 s.

Using observations in different spectral windows we can track the wave propagation through phase difference of oscillations at different heights, which allows an opportunity to estimate the energy flux carried by the waves.

## Observational requirements

- **Target:** Chromospheric network near disk center.
- **Antenna configuration:** C40-2
- **Pointing/FOV:** single *(FOV: maximum possible)*
- **Receiver**: One run in band 3 (100 GHz), one run in band 6 (230 GHz).
- **Temporal resolution:** 2 s
- **Duration:** 10 min

## References

*Katsiyannis, A. C., Williams, D. R., McAteer, R. T. J., Gallagher, P. T., Keenan, F. P. and Murtagh, F., 2003, A&A, 406, 709*

*Jess, D. B., Andić, A., Mathioudakis, M., Bloomfield, D. S. and Keenan, F. P., 2007, A&A, 473, 943*

*Morton, R. J., Verth, G., Hillier, A. and Erdélyi, R., 2014, ApJ, 784, 29*

*Williams, D. R., Mathioudakis, M., Gallagher, P. T., Phillips, K. J. H., McAteer, R. T. J., Keenan, F. P., Rudawy, P. and Katsiyannis, A. C., 2002, MNRAS, 336, 747*

*de Wijn, A. G., Rutten, R. J. and Tarbell, T. D., 2005, A&A, 430, 1119*

*Wunnenberg, M., Kneer, F. and Hirzberger, J., 2002, A&A, 395, L51*

*Zaqarashvili, T. V., Khutsishvili, E., Kukhianidze, V. and Ramishvili, G., 2007, A&A, 474, 62*### G2.6 - Science Case SC-G6: Seismology of chromospheric fibrils

**Key questions: Do chromospheric fibrils support quasi-periodic propagating waves? If detected, can the spatial and/or time signatures of these waves be used for the seismology of chromospheric fibrils?**

Dense magnetic flux tubes can efficiently guide magnetohydrodynamic fast waves (Edwin & Roberts 1988). A sudden, localised disturbance acting on a dense flux tube creates a wave pulse that propagates along the tube's length. Because of the dispersive nature of the tube, the initial pulse evolves into a wave train that causes quasi-periodic oscillations (e.g. Edwin & Roberts 1988) of the plasma parameters and the magnetic field. The structure of these wave trains strongly depends on the ratio of the tube to the environment density (Nakariakov et al. 2004) and so their observation has robust potential in the seismology of dense magnetic tubes. Quasi-periodic propagating waves have been observed as coronal intensity variations in white

http://ssalmon.uio.no

light (Williams et al. 2001) and at extreme ultraviolet wavelengths (e.g. Nisticò et al. 2014). The detection of propagating wave trains caused by a sudden compression or expansion of a chromospheric fibril is the specific aim of this proposal. The involved temporal variations have periods ~2 $\tau_A$ and last for ~40 $\tau_A$ (where $\tau_A$ is the ratio of tube radius, R, to Alfvén speed, $v_A$). The wave train contains wavelengths ~4R and total length ~15R. Fibrils with diameter ~1 arcsec and $v_A$=50-100 km/s can guide wave trains with periods ~3.5-7 s and wavelengths ~2 arcsec. In addition, expected temperature variations of these events are of the order of a few percent the unperturbed temperature.

**Expected achievement:** determination of density contrast of chromospheric fibrils.

**Observational requirements**

- **Target:** chromospheric fibrils in the vicinity of a pore or a non-flaring active region.
    - **Observable signatures with ALMA:** quasi-periodic temperature variations with periods around 5s and duration around 100s.
- **Array configuration:** Interferometric mode, C40-2 (or C40-3 if available).
- **Pointing/FOV:** single.
- **Receiver:** band 6.
- **Temporal resolution:** 2 s.
- **Duration:** 30 min.

**Possible issues**

- Absence of the target events because of lack of proper excitation.
- Target events evolving into noisy signal because of interference of multiple pulses launched in rapid succession.
- Although the spatial resolution of ALMA may be insufficient to resolve the finer individual structures, a quasi-periodic wave train signal may stand out in front of the background and still be detectable.

**References**

*Edwin, P. M., Roberts, B. 1988, A&A, 192, 343*
*Nakariakov, V. M. et al. 2004, MNRAS, 349, 705*
*Nisticò, G., Pascoe, D. J., Nakariakov, V. M. 2014, A&A, 569, A12*
*Williams et al. 2001, MNRAS, 326, 428*

### G2.7 - Science Case SC-G7: Short period oscillations in filament threads

**Key Questions: Detection of very short period oscillations in filament threads.**

To study sub-arcsecond solar structures such as filament threads (~ 0.3 arcsecond), observations with high quality, high spatial and high temporal resolution are required, which resolve individual threads and allow us to perform seismological studies. The reported oscillatory periods in small amplitude oscillations have been classified in four groups (Lin, 2011): very short periods (≤ 1 min), short periods (1 min – 20 min), intermediate (20 - 40 min) and long (40 - 100 min). Taking into account the short lifetime (~ 10 – 15 min) of filament threads, we should concentrate on the detection of very short period oscillations. On the other hand, the damping of small amplitude oscillations in short spatial and temporal scales is also well known (Arregui & Ballester, 2011) and should be another observational target. From the theoretical point of view, since these oscillations have been interpreted in terms of MHD waves, the relevance of each potential damping mechanism can be assessed by comparing the spatial and temporal scales produced by each of them with those obtained from observations.

**Observational requirements**

- **Target:** Solar filaments near disk center; interferometric mode
- **Antenna configuration:** C40-3
- **Pointing/FOV:** single, sit-and-stare.
- **Receiver:** Band 6 (230 GHz)
- **Temporal resolution:** 2 s
- **Duration:** 10 min

**Possible issues**

- Absence of filaments near disk center.
- **Alternative target:** Vertical fine structures of prominences on the limb

**References**

*Arregui, I. & Ballester, J. L. 2011, Space Science Reviews, 158, 169*
*Lin, Y. 2011, Space Science Reviews, 158, 237*

### G2.8 - Science Case SC-G8: Quasi-periodic pulsations (QPPs)

**Key Questions: Fine spatial and time structure of phases, periods and power of QPP in flares**

Light curves of the energy releases in solar and stellar flares have quasi-periodic pulsations (QPP, see Nakariakov & Melnikov 2009 for a review; and Kupriyanova et al. 2010 for demonstration of the ubiquity of this phenomenon). QPP range form a fraction of a second up to tens of minutes, and are seen, often simultaneously, in all observational bands, from radio to gamma-rays. QPP are seen in a broad range of flares, from the low C- to high X-class (see http://goo.gl/iylBnM for a catalogue of published works on QPP). Physical mechanisms for QPP are still debated, but as in many cases QPP appear in the signals produced by non-thermal electrons, it is natural to associate QPP with spontaneous (e.g. Kliem et al. 2000; Tajima et al. 1987) or induced (e.g. Chen & Priest 2006; Nakariakov et al. 2006) periodic regimes of magnetic reconnection. The QPP observed at the lower layers of the solar atmosphere are thus produced by e.g. some periodic variation of the particle acceleration rate or some spatial non-uniformity of the acceleration region. Therefore, any model of flaring energy release could not pretend to be complete without explaining QPP. Moreover, as QPP detected in stellar superflares have similar properties as solar QPP, there is a question whether they are caused by the same physical mechanisms, and hence are directly linked to the estimation of the probability the Sun producing a highly-geoeffective superflare (Anfinogentov et al. 2013) and conditions for it.

One of the crucial obstacles in the observational study of QPP is the lack of simultaneously high spatial and time resolution. ALMA's combination of high spatial and time resolution and sensitivity would be highly useful in establishing the fine spatial structure of the oscillatory flaring source, including the distribution of the oscillation phase, period and power. This information will allow us to discriminate between different models of the QPP production. Another important information is detection of multi-periodic QPP. Several examples have already been detected in solar (e.g. Van Doorsselaere et al. 2011) and stellar flares (Pugh et al. 2015). The very presence of the multiple periodicities is a strong evidence in favour of MHD-wave-based mechanisms for QPP. Moreover, the period ratio provides us with invaluable seismological information about the flaring sites and their surroundings.

As solar flares appear suddenly, it is quite likely that the data necessary for this project will be obtained as by-products of other observational projects. Another way would be to perform a sit-and-stare observation of a candidate active region. Many QPP studies have used sun-as-a-star observations. This is possible since only a tiny part of the solar atmosphere is highlighted by the flare, and thus providing a natural fine resolution. Thus, this science case could also be investigated with single-dish observations dish with reduced spatial resolution but increased time resolution.

**Observational requirements**

- **Target:** Footpoints of an active region arcade.
- **Antenna configuration:** C40-1
- **Pointing/FOV:** single. The FOV should cover a sufficient fraction of the active region, ideally including both footpoints of an active region magnetic structure, on both sides from the neutral line.
- **Receiver:** Band 6 (230 GHz)
- **Temporal resolution:** 2 s or better if possible.
- **Duration:** 30 min *(plus calibration interruptions; those as short as possible)*

**References**

*Anfinogentov, S., et al. 2013, Astrophys. J. 773, 156*
*Chen, P. F., & Priest, E. R. 2006, Solar Phys. 238, 313*
*Kliem, B., et al. 2000, Astron. Astrophys. 360, 715*
*Kupriyanova, E. G., et al. 2010, Solar Phys. 267, 329*
*Nakariakov, V. M., et al. 2006, Astron. Astrophys. 452, 343*
*Nakariakov, V. M., & Melnikov, V. F. 2009, Space Sci. Rev. 149, 119*
*Pugh, C., et al. 2015, Astrophys. J. Lett. 813, L5*
*Tajima, T., et al. 1987, Astrophys. J. 321, 1031*
*Van Doorsselaere, T., et al. 2011, Astrophys. J. 740, 90*

## G2.9 - Science Case SC-G9: Sunspot oscillations

**Key Questions: Fine spatial structure of umbral oscillations**

Oscillations in sunspot umbrae are one of the most pronounced wave phenomena in the solar atmosphere (Solanki 2003). At the chromospheric heights the dominating period of the oscillations is in the vicinity of 3 minutes, while the spatial distribution of the periods and spectral power has clear radial structuring (Reznikova & Shibasaki, 2012) with longer period oscillations situated at a larger radial distance from the umbra centre. It is commonly accepted that chromospheric 3-min oscillations are associated with slow magnetoacoustic waves, and that their period is prescribed by the plasma temperature (Bogdan 2000). Despite significant progress in observational studies and theoretical modelling, there are a lot of open questions connected with 3-min oscillations.
Although theory explains well the radial dependence of the oscillation period, linking it with the magnetic field inclination, it is not clear what determines the fine spectral structuring of umbral oscillations. In particular, the reasons for the appearance of distinct multiple peaks in the power spectrum are not known. The spatial structure of the oscillation phase, including the transverse extend of coherent oscillations remains a puzzle. Recent finding (e.g. Sych & Nakariakov 2014) show that, while the oscillation period gradually increases with the distance from the umbra centre, the power changes abruptly. Moreover, the enhancements of the oscillation power, or wave fronts, are found to have a distinct structure consisting of an evolving two-armed, gradually rotating spiral and a stationary circular patch at the spiral origin, situated near the umbra centre. This behaviour attracts attention as it may shed light on the sub-photospheric structure of the sunspot magnetic field, which may provide us with important clues about sunspot formation mechanisms. In addition, there is evidence that various details inside the umbra, such as light bridges show the oscillatory behaviour different from umbral (Yuan et al. 2014), which allows one to perform seismological probing of these details. Also, as a side result we may expect to get some evidence of fast magnetoacoustic waves in sunspot umbra and their coupling with slow modes (e.g. Felipe et al. 2010).

We aim utilise ALMA's enhanced spatial resolution and sensitivity to reveal the fine spatial structure of the phase and power of 3-min oscillations, and get insight in the subphotospheric structure of sunspot's magnetic field. As umbral oscillations are always present in sunspots, the observational programme would be to choose a field of view around a sunspot or a pore, and observe it in the seat-and-stare mode for a long time. There is no need for a high time resolution, and we believe that a 10 s cadence time would be sufficient for a detailed study of 3-min oscillations.

**Observational requirements**

- **Target:** Footpoints of an active region arcade.
- **Antenna configuration:** C40-3 in band 6 (230 GHz), or, if C40-3 is not available, C40-2.
- **Pointing/FOV:** single.
- **Receiver:** Band 6 (230 GHz) (and, if possible, also one run in Band 3 (100 GHz)). Please note that, for this science case, it would be desirable to observe in bands 3 and 6 simultaneously, which may only become possible by using sub-arrays in Cycle 5 or possibly once switching receiver bands takes less than 10 s (if possible).
- **Temporal resolution:** 10 s
- Duration: 30 min (plus calibration interruptions; those as short as possible)
- **FOV:** a sufficient fraction of the active region, ideally including both footpoints of an active region magnetic structure, on both sides from the neutral line.

## 15. CHROMOSPHERIC AND CORONAL HEATING (TEAM J)

**List of contributing team members:**
P. Antolin, T. Ayres, M. Carlsson, G. Doyle, M. Loukitcheva, S. Tiwari, S. Wedemeyer

**Lead author:** S. Wedemeyer *(sven.wedemeyer@astro.uio.no)*

### J1 - SUMMARY OF SCIENCE CASES AND KEY QUESTIONS

**Science Case SC-J1: Chromospheric to coronal plasma response**

- **Key Question SC-J1-KQ1:** How does the chromospheric plasma influence the adjacent coronal plasma?

**Science Case SC-J2: Dynamic nature of atmospheric heating**

- **Key Question SC-J2-KQ1:** Is the heating of the chromospheric and coronal plasma in quiet Sun regions mostly continuous or rather episodic?
- **Key Question SC-J2-KQ2:** Is the heating of the chromospheric and coronal plasma in active regions mostly continuous or rather episodic?

**Science Case SC-J3: Penumbral jets and bright dots**

- **Key Question SC-J3-KQ1:** How are penumbral jets and bright dots contributing to the heating of the solar atmosphere?

**Science Case SC-J4: Thermal inhomogeneity of the solar corona**

- **Key Question SC-J4-KQ1:** What is the role of thermal instability in the solar corona?
- **Key Question SC-J4-KQ2:** Is coronal heating a drastically inhomogeneous process with an extremely cold and frequent counterpart?
- **Key Question SC-J4-KQ3:** What is the role of coronal rain, the direct consequence of thermal instability, in the chromosphere-corona mass cycle?
- **Key Question SC-J4-KQ4:** What is the degree of spatial inhomogeneity in the solar corona? Does thermal instability play an essential role in defining spatial inhomogeneity such as fundamental strands and turbulence?

### J2 - DETAILED DESCRIPTIONS OF SCIENCE CASE

#### J2.1 - Science Case SC-J1: Chromospheric to coronal plasma response

**SC-J1-KQ1: How does the chromospheric plasma influence the adjacent coronal plasma?**

The corona and chromosphere constitute a closely interacting plasma system in which perturbations in one can strongly affect the other. This is the case for the unknown coronal heating mechanism. Due to the very efficient thermal conduction in the corona, a heating event in the corona unavoidably transfers energy into the chromosphere, which leads to heating and chromospheric evaporation. Moreover, due to the strong radiative losses in the chromosphere, the heating events there must be a few orders of magnitude more frequent and effective than in the corona. Investigating the close relation between the chromosphere and the corona is thus an important approach to understanding the heating mechanisms in the solar atmosphere.

A heating mechanism in a coronal structure such as a loop entails observable consequences in the chromospheric part of the same magnetic field structure (its footpoint). Thus, by observing the footpoints of loops in ALMA chromospheric passbands we can capture the chromospheric response to heating events along the loop (wherever these may be). These events will manifest as increases in the temperature brightness and optical thickness (chromospheric evaporation, even if small, leads to an increase in the geometrical thickness for vertical line-of-sights centred at the loop footpoints).

By co-observing with transition region (TR) and coronal instruments such as IRIS and SDO/AIA we can then correlate such heating events with intensity brightening and line broadening events in the same magnetic field structure. SDO/HMI can provide accurate magnetograms through which magnetic field extrapolations can be done. Such magnetic field maps will ensure that we are observing the same magnetic field structure with the multiple instruments. The interpretation of the observations can be supported by numerical modelling, e.g. with the RADYN code.

**Observational requirements**

- **Target:** A network patch where coronal loops are rooted (best in an active region for loop visibility in AIA; AIA may actually serve as guide previous to observations).
- **Array configuration:** Compact (C40-1-C40-3).
- **Pointing/FOV:** Single point interferometry (sit-and-stare, 58" FOV).
- **Receiver:** Band 3.
- **Temporal resolution:** Heating events in the chromosphere are expected to range from a few seconds up to hundreds of seconds. Thus, a high cadence is desired.
- **Duration:** 1 hour continuous observing (with calibration breaks being as short as possible).
- **Co-observation** with IRIS, Hinode/XRT and EIS and SDO/AIA & HMI

## J2.2 - Science Case SC-J2: Dynamic nature of atmospheric heating

**Key Questions: SC-J2-KQ1:** Is the heating of the chromospheric and coronal plasma in quiet Sun regions mostly continuous or rather episodic?

**SC-J2-KQ2:** Is the heating of the chromospheric and coronal plasma in active regions mostly continuous or rather episodic?

Today many potential mechanisms are known which could provide enough energy to explain the heating of the chromosphere and corona. On the other hand, it is still unclear which mechanisms dominate, how much they contribute to the heating and how this differs between regions of different activity level. A first step on the way to entangle the different heating mechanisms is to ask whether chromospheric and coronal heating is continuous or rather consists of sequences of episodic heating events.

Heating events should be detectable by changes of the chromospheric gas temperature, which can be derived from ALMA observations. Co-observations with IRIS would add complementary information on the changes of the chromospheric plasma state. The measured rates of the temperature changes and temporal changes in velocities and spectral line properties will give first ideas how episodic the chromospheric heating is. The same study can be carried out in quiet Sun regions and active regions.

**Observational requirements**

- **Target/FOV:**
  1. Quiet Sun
  2. Active Region
- **Array configuration:** Compact (C40-1-C40-3).
- **Pointing/FOV:** Single, sit-and-stare.
- **Receiver:** Band 3, making use of all 4 sub-bands.
- **Temporal resolution:** As high as possible, i.e. 2s in Cycle 4.
- **Duration:** One hour quiet Sun plus one hour for an active region.
- **Co-observing:** IRIS.

**Technical considerations**

What would be lowest cadence that would still make sense for the scientific objective? Integrating over time will not improve much the restored resolution but will smear out changes on the shortest timescales which are usually connected to the smallest spatial scales. On the other hand, integrating over longer times would result in smaller flux uncertainties, which means that brightness changes with smaller amplitudes could be restored. However, the optimum integration should not exceed 30s-1min to enable the detection of episodic heating events.

## J2.3 - Science Case SC-J3: Penumbral jets and bright dots

**SC-J3-KQ1:** How are penumbral jets and bright dots contributing to the heating of the solar atmosphere?

Penumbral jets (PJs) are transient narrow bright events constantly occurring in sunspot penumbrae. They were first observed and characterized by Katsukawa et al. (2007) by using chromospheric CaII H-line images obtained with Hinode (SOT/FG).

The estimated thermal energy ($3/2$ n k T V) of a typical jet is of the order of that of a nanoflare ($10^{23}$ erg). Thus, PJs could potentially play an important role in chromospheric/TR/coronal heating above sunspots. The formation mechanism and exact location of PJs on the photosphere, their contribution to chromospheric heating of sunspot penumbra and response to the transition region (TR) and corona above sunspots still remain elusive.

Moving bright penumbral dots (BDs) may form as a result of reconnection between two inclined fields or as a result of downflowing plasma hitting the high density in the lower atmosphere. BDs have much longer lifetimes (1-5 minutes) than PJs (less than a minute) but some of the short lived BDs might form by the same mechanism as PJs. Both have enough thermal energy to show up in the TR and corona.

**Key idea:** It has been proposed recently that normal PJs (150 - 400 km wide) might form by magnetic reconnection between spine field and opposite polarity field along the lateral sides of penumbral filaments, whereas larger repetitive jets (>400 km wide) form at tails of penumbral filaments by reconnection between opposite polarity fields of spines and filament tails. The observational evidence of this proposal remains missing owing to high cadence and high spatial resolution data of a sunspot observed simultaneously in the photosphere and chromosphere. The high resolution chromospheric observations of ALMA and a simultaneous observation of Hinode (FG) will allow to track the feet of the normal and larger PJs and verify their formation mechanism.

Although PJs are narrow (150-600 km) transient events, they are abundant, and might be responsible for heating chromosphere/ TR/corona above sunspots. Some of the larger PJs directly show their signatures in the TR and coronal wavelengths. Simultaneous observations of a sunspot penumbra by ALMA, IRIS and AIA will allow us to follow these jets in different atmospheric heights and temperatures and quantify their contribution to chromospheric/TR/coronal heating.

Both of the above is also true for BDs, i.e., how and where BDs form and whether they significantly heat the chromosphere/TR/corona of sunspots.

**Observational requirements**

To spatially track the locations of these chromospheric/TR events on the photosphere, we need high resolution co-observations of a sunspot with Hinode (SOT/FG), and IRIS. AIA will be available to guide and complement these data sets.

- **Target:** High-resolution interferometric imaging following a developed sunspot preferably close to solar disk center.
- **Array configuration:** Compact (C40-1 - C40-3).
- **Pointing/FOV:** Single (50 - 60 arcsec FOV).
- **Receiver:** Band 3.
- **Temporal resolution:** A cadence of 10 - 20 s cadence should be sufficient.
- **Duration:** One hour.
- **Co-observing:** Hinode (SOT/FG), IRIS, SDO/AIA.

## J2.4 - Science Case SC-J4: Thermal inhomogeneity of the solar corona

**Key Questions: SC-J4-KQ1: What is the role of thermal instability in the solar corona?**

**SC-J4-KQ2: Is coronal heating a drastically inhomogeneous process with an extremely cold and frequent counterpart?**

**SC-J4-KQ3: What is the role of coronal rain, the direct consequence of thermal instability, in the chromosphere-corona mass cycle?**

**SC-J4-KQ4: What is the degree of spatial inhomogeneity in the solar corona? Does thermal instability play an essential role in defining spatial inhomogeneity such as fundamental strands and turbulence?**

A major question in coronal heating concerns the spatial distribution of the heating mechanism in loops. A large amount of evidence exists suggesting that this spatial distribution is mostly footpoint concentrated in active region (AR) loops, implying that thermal conduction plays a major role in transporting the energy from energetic events at the footpoint of loops into the corona (Hansteen et al. 2015, De Pontieu et al. 2011). Such spatial distribution of the heating in loops combined with the radiative properties of plasmas leads to coronae that are non-uniformly heated and to states of thermal non-equilibrium in which the coronae are thermally unstable and therefore strongly dynamic (Parker 1953, Field 1965, Goldsmith 1971). A direct consequence of this are limit cycles of hot and cool stages in the thermal evolution of active region loops (condensation - evaporation cycles) (Mok, Y. et al. 1990, Antiochos et al. 1999, Froment et al. 2015). In the cool stage of these loops coronal rain is produced, that is, cool and dense condensations spontaneously occurring in the corona (Leroy 1972, Schrijver 2001, Antolin, P. et al. 2010). These condensations have a wide range of temperatures, from very cool (2000 K) to the transition region (TR) range (100000 K) (Antolin et al. 2015, Scullion, E. et al. 2014). Recent studies suggest that thermal instability is a common phenomenon, and the coronal rain that results is often very cold (Antolin et al. 2012a, 2012b, 2015, Fang et al. 2015). This suggests that coronal heating is an extremely inhomogeneous process, with a very cold and important counterpart.

By detecting coronal rain in active region coronae with ALMA and by further measuring the temperature in coronal rain we can gain insight into the thermal instability and coronal heating mechanisms in the solar atmosphere. Furthermore, since the partially ionised material that composes coronal rain is still strongly bound to the magnetic field, the rain serves as high resolution tracers of the magnetic field topology. Thus, by following coronal rain in multiple instruments such as Hinode/SOT, IRIS and AIA, covering the chromospheric to coronal temperature range, we can map at high resolution the spatial inhomogeneity of the thermal structure of these loops.

At chromospheric temperatures coronal rain is observed to be clumpy, with width distributions that peak at increasingly small values (less than 0.3") for higher resolution, and more scattered length distributions that can be as long as a few tens of arcseconds. Clumps do not occur individually but are often observed as belonging to "showers", that is myriads of clumps falling simultaneously over significant periods of time (around 10 min or more, and repeatedly) in a single loop structure that generally spans a width of a few arcseconds. Plasma conditions inside a rain clump can present large inhomogeneities, similar to those of prominences, with average temperatures and densities around 5000 K and $10^{11}$ cm$^{-3}$, respectively, in the core, surrounded by a more diffuse envelope with higher temperatures and lower densities more typical of the transition region (Antolin et al. 2015). At the resolution of band 3 and band 6 of ALMA we expect to observe rain showers (and large rain clumps), for which the brightness temperatures and optical thickness have been estimated in a similar way as for prominences (Heinzel et al. 2015), and are provided in the following table (assuming for the shower a filling factor of 0.5 and a width of 2 Mm):

**Shower**
- Brightness temperature extrema: @100 GHz: [450, 90000] K, @230 GHz: [20, 40000] K
- Brightness temperature average: @100 GHz: 5000 K (for an average density a of 5*10^10 cm^-3 and average temperature of 8000 K) @230 GHz: 2200 K
- Optical thickness extrema: @100 GHz: [0.05, 45], @230 GHz: [0.025, 20]
- Optical thickness average: @100 GHz: 0.625, @230 GHz: 0.27

Since the rain cannot only be observed off-limb but also on-disc, and especially above sunspots, where the background chromospheric intensity is low and the contrast of the rain is high (Scullion et al. 2014, Antolin, P. et al. 2012a, 2012b, Ahn et al. 2014, Kleint et al. 2014 ), we propose the following observational setup:

### Observational requirements

- **Target:** Sunspot close to solar disk center.
- **Array configuration:** Compact (C40-1-C40-3).
- **Pointing/FOV:** Single point interferometry (sit-and-stare).
- Band 3: Centered on a sunspot on-disc not so far from solar disk center (58" FOV).
- Band 6: Centred on a sunspot umbra (SDO/AIA can serve as guide for selecting the FOV center), (25" FOV).
- **Receiver:** The observation sequence should be carried out first with band 3, then with band 6.
- **Temporal resolution:** A cadence of 10 seconds is desired in order to be able to follow large clumps and showers as they fall into the sunspot umbra, in ALMA and the accompanying co-observing instruments.
- **Duration:** One hour continuous observing (plus calibration).
- **Context observations:** The H-alpha ground-based instruments that have been so far proposed + IRIS + Hinode

## 16. LIMB-BRIGHTENING STUDIES (TEAM P)

**List of contributing team members:**

P. Antolin, J. Costa, G. Gimenez de Castro, H. Hudson, R. Oliver, C. Selhorst, S. Wedemeyer

**Lead author**: H. Hudson (hhudson@ssl.berkeley.edu)

### P1 - SUMMARY OF SCIENCE CASES AND KEY QUESTIONS

**Science Case SC-P1: The mean vertical structure of the solar atmosphere**

- **Key question SC-P1-KQ1:** How does the global magnetic field affect the structure?
- **Key question SC-P1-KQ2:** How does the magnetic network affect the vertical structure?

**Science Case SC-P2: Variation of the brightness temperature on network and granule scales**

- **Key question SC-P2-KQ1:** How does the brightness temperature vary on network and granule scales?

### P2 - DETAILED DESCRIPTIONS OF SCIENCE CASES

#### P2.1 Science Case SC-P1: The mean vertical structure of the solar atmosphere

**Key questions: SC-P1-KQ1: How does the global magnetic field affect the structure?**
**SC-P1-KQ2: How does the network affect the vertical structure?**

The general center-to-limb variation of brightness in the ALMA spectral range directly reflects the vertical structure of the physical temperature of the medium. This reflects itself in the absolute continuum intensity inferred from disk-center observations, and its variation through the range of viewing angles. Such center-to-limb observations could be used to produce a series of 1D model atmospheres that accurately reproduce the observed continuum. Such a set of models should provide the best possible interface for full dynamical modeling of the lower atmosphere and global tests for the physical assumptions necessary for the modeling work.

Essentially Team P aims at the eventual construction of mean models of the atmosphere, at the resolution permitted by ALMA interferometry. The limb-darkening function in its dependence on wavelength is the key measureable parameter, and its interpretation depends at the simplest level only on the continuum opacity model. That said, many factors can intervene at the level of precision anticipated for ALMA. These include both the physics (coronal rain, for example, or unexpected bound-bound opacities) and the technology (how well can we understand the calibrations of beam profile, radiometry, and interferometry?). Similar efforts have happened historically (e.g. Ewell et al. 1993, Bastian et al. 1993) but ALMA will enable now a much higher precision.

The basic observational material for limb-darkening research strongly overlaps with that described by other SSALMON expert teams, notably team K ("Quiet Sun regions") but probably also others (Team O: "Implications for stellar physics"). The basic observational material has two distinct parts:

**1) Single-dish (SC-P1-KQ1).** The ALMA dishes and their calibrations can be expected to be the very finest ever made, with superb advantages in many ways. The observational problem in Cycle 4 will be to explore at as fine a level as possible the beam patterns and the inferences that can be made about center-to-limb variation, including absolute central temperature, in Bands 3 and 6.

**2) Interferometric (SC-P1-KQ2).** The ALMA interferometry will characterize the spatial structure; for example, with the Cycle 4 bands 3 and 6 we will be able to distinguish network cells precisely at disk center.

**Observational requirements**

- **Observing setup:** The SWEEP set-up described in Table 6.2 would be ideal as it offers a sequence with many pointings without the need for long sequences at the individual pointings (because we are interested in the mean structure).
- **Array configuration:**
  - Interferometric array in a compact configuration without mosaicking but individual pointings (SWEEP mode) plus fast scanning.
  - Fast scanning with 3-4 TP antennas as available. The TP antennas should perform full-Sun FOV scans in double-circle configuration (2400" diameter).
- **Pointing/scan configuration:** SWEEP sequences of positions for interferometry at equal steps of FWHM/2 from disk center to 1.1 R_sun, which is about 1060". For Band 3, this requires about 35 pointings, thus 3.5 SWEEP blocks with 10 pointings each. For band 6, this would require 8 blocks with a total of 80 pointings. In the long run, multiple radial scans on heliographic N, E, S, W are preferred. For Cycle 4, we propose to start with a scan from disk center to solar north pole in band 3 and to repeat the scan in band 6.
- **Receiver:** Full sequence first in band 3, then in band 6.
- **Temporal resolution:** Not important. For each pointing any choice from 2s to max. 60s is possible.
- **Duration:** The interferometric SWEEP sequence in Band 3 takes 3.5x 10min + 4x3min calibration (47min in total); the sequence in Band 6 takes 8x(10min+3min), thus 104min in total . Including all calibrations, the program would take 211 min for completion.

- **Context observations:** Full disk maps from SDO will be useful for comparison.

**References:**

*Ewell, M.W. et al. 1993, ApJ 403, 426*
*Bastian, T.S. et al. 1993, ApJ 415, 364*

**Please also see the following science cases, which investigate the center-to-limb variation:**
- **SC-K1:** The brightness temperature distribution in quiet Sun regions and its center-to-limb variation
- **SC-K2:** Local slopes of the brightness temperature spectrum

**P2.2 Science Case SC-P2: Variation of the brightness temperature on network and granule scales**

**Key question SC-P2-KQ1: How does the brightness temperature vary on network and granule scales?**

What is the nature of the "true" body of the Sun, i.e. that unperturbed by magnetic concentrations? Such a question could not have been answered prior to high-resolution observations of the true continuum at disk center, which ALMA offers us. With this measurement we will be able to understand what fraction of the solar luminosity resides in mainly thermal effects, rather than in wave or magnetic activity creating emission excesses in at extreme wavelengths (mainly in the UV energetically, and of course signatures of magnetic activity in the infrared and radio wavelengths also arise in this excess.

The basic measurements will come from those described above, but at disk center. The measurement should be repeated each year in order to understand the solar-cycle effects.

**Observational requirements**

Same as for SC-P1. See Sect. P2.1 above.

**References:**

*Ewell, M.W. et al. 1993, ApJ 403, 426*
*Bastian, T.S. et al. 1993, ApJ 415, 364*

**Please also see the following science cases, which investigate the center-to-limb variation:**
- **SC-K1:** The brightness temperature distribution in quiet Sun regions and its center-to-limb variation
- **SC-K2:** Local slopes of the brightness temperature spectrum